\documentclass[
preprint,
superscriptaddress, 
aps,
pra, 
onecolumn, 
notitlepage,
amsmath, 
longbibliography,
amssymb
]{revtex4-1}

\usepackage{graphicx}
\usepackage{bm}
\usepackage{color}
\usepackage[svgnames]{xcolor}
\usepackage{hyperref}
\usepackage[mathlines]{lineno}
\usepackage{epstopdf}
\usepackage{stmaryrd}
\usepackage{ulem} 

\usepackage{algorithm}
\usepackage{algpseudocode}
\usepackage{enumitem}
\newcommand{\mc}{\mathcal}
\newcommand\Algphase[1]{%
	\vspace*{-.7\baselineskip}\Statex\hspace*{\dimexpr-\algorithmicindent-2pt\relax}\rule{\textwidth}{0.4pt}%
	\Statex\hspace*{-\algorithmicindent}\textbf{#1}%
	\vspace*{-.7\baselineskip}\Statex\hspace*{\dimexpr-\algorithmicindent-2pt\relax}\rule{\textwidth}{0.4pt}%
}
\def\NFCA{NaFe$_{1-x}$Co$_{x}$As }

\hypersetup{
pdfpagemode={UseOutlines},
pdfpagelayout={SinglePage},
bookmarksopen=true,
bookmarksnumbered = true,
pdfstartview={FitV},
colorlinks = true,
linkcolor = DarkBlue,
citecolor = DarkBlue,
urlcolor = DarkBlue
}

\DeclareMathOperator*{\argmin}{arg\,min}
\newcommand*\diff{\mathop{}\!\mathrm{d}}
\newcommand{\ket}[1]{\left| #1 \right>} 
\newcommand{\bra}[1]{\left< #1 \right|} 
\renewcommand{\Im}{\operatorname{Im}}

\def\NFA{NaFeAs}

\begin{document}

\title{Dictionary Learning in Fourier Transform Scanning Tunneling Spectroscopy}

\author{Sky C. Cheung \footnote{These authors contributed equally to this work}}
\affiliation{Department of Physics, Columbia University, New York, NY 10027 USA}
\author{John Y. Shin \footnotemark[1]}
\affiliation{Department of Physics, Columbia University, New York, NY 10027 USA}
\author{Yenson Lau}
\affiliation{Department of Electrical Engineering, Columbia University, New York, NY 10027 USA}
\author{Zhengyu Chen}
\affiliation{Department of Electrical Engineering, Columbia University, New York, NY 10027 USA}
\author{Ju Sun}
\affiliation{Department of Electrical Engineering, Columbia University, New York, NY 10027 USA}
\author{Yuqian Zhang}
\affiliation{Department of Electrical Engineering, Columbia University, New York, NY 10027 USA}
\author{John N. Wright}
\email{johnwright@ee.columbia.edu}
\affiliation{Department of Electrical Engineering, Columbia University, New York, NY 10027 USA}
\author{Abhay N. Pasupathy}
\email{pasupathy@phys.columbia.edu}
\affiliation{Department of Physics, Columbia University, New York, NY 10027 USA}

\date{\today}



\maketitle

\textbf{
	Modern high-resolution microscopes, such as the scanning tunneling microscope, are commonly used to
	study specimens that have dense and aperiodic spatial structure~\cite{Crommie1993, Hoffman2002,
		Stroscio2007, RoushanSeo_Nature_2009, Hanaguri2010, Rosenthal2014, Arguello2015}. Extracting meaningful
	information from images obtained from such microscopes remains a formidable
	challenge~\cite{WangLee_PRB_2003, Heller2003, Kivelson2003}. Fourier analysis is commonly used to analyze
	the underlying structure of fundamental motifs present in an image~\cite{Mcelroy2001, Hoffman2002,
		WangLee_PRB_2003, RoushanSeo_Nature_2009, Rosenthal2014, Arguello2015}. However, the Fourier transform fundamentally
	suffers from severe phase noise when applied to aperiodic images~\cite{Pillow2013_Neurology,
		Faghih2014_Biology}. Here, we report the development of a new algorithm based on nonconvex
	optimization, applicable to any microscopy modality, that directly uncovers the fundamental motifs
	present in a real-space image. Apart from being quantitatively superior to traditional Fourier
	analysis, we show that this novel algorithm also uncovers phase sensitive information about the
	underlying motif structure. We demonstrate its usefulness by studying scanning tunneling microscopy
	images of a Co-doped iron arsenide superconductor and prove that the application of the
	algorithm allows for the complete recovery of quasiparticle interference in this material. Our phase sensitive quasiparticle interference imaging results indicate that the pairing symmetry in optimally doped NaFeAs is consistent with a sign-changing $s^{\pm}$ order parameter.
}

The past few decades have seen dramatic advances in the understanding of the structure of materials
via scattering and microscopy techniques. Scattering techniques are useful when perfect periodicity
exists in a material, while microscopy is well-suited for specimens that lack periodicity. Recent
advances in microscopy techniques, when coupled with improved computing power, have enabled the
scientific community to generate massive, multi-dimensional spatial images of specimens as a
function of control parameters such as time, energy, and applied stimulus. Examples of such advanced
tools include super-resolution optical microscopy to inspect the structure of proteins beyond the
diffraction limit~\cite{Betzig2006_PALM}, scanning transmission electron microscopy to examine the
chemical structure of materials at the atomic scale~\cite{Muller2009}, and scanning tunneling
microscopy (STM) to visualize the quantum electronic structure of surfaces with atomic
resolution~\cite{BinnigRohrer1983}. Fundamentally, a microscope image represents the interaction
between the probe and the specimen, and often times sophisticated analysis must be performed to
uncover the scientific content present in the image. Specimens of interest for STM studies include
metals~\cite{Crommie1993}, two-dimensional materials~\cite{Stroscio2007}, unconventional
superconductors~\cite{Hoffman2002}, topological materials~\cite{RoushanSeo_Nature_2009,
	Wiesendanger2015_Skyrmion}, and charge~\cite{Arguello2015} and spin~\cite{Hanaguri2010, Wahl2014}
ordered materials, among others. Image analysis of these materials has provided several unique
insights into the quantum electronic structure and interactions present within them.  Many
microscopy techniques utilize the Fourier Transform (FT)~\cite{Kimoto201231_TEM_FT,
	Jaffe1987_HREM_FT} for analysis, revealing the characteristic wavelengths present in the image,
which are then related to a scientific theory of the specimen being studied. When perfect
periodicity exists in an image, the FT provides a concise and accurate description of the image.
However, when applied to aperiodic images, the FT suffers from phase noise leading to a fundamental
loss of information. With the proliferation of new computing techniques, one may wonder if the
maturation of optimization algorithms can be leveraged to extract more information from a microscopy
image than through the FT. In this work, we consider a class of images that are of particular
importance to microscopy -- those that can be perceived as a basic motif, called a kernel, that is
repeated aperiodically across the image. Examples of kernels include electronic scattering patterns
around atomic defects (in STM) and fluorescence from individual proteins (in optical microscopy). We
present the development of an algorithm, based on nonconvex optimization, for analyzing such images
that quantitatively extracts the principal motifs present in an image without estimating or
averaging over instances of the defect. We demonstrate that this algorithm can elucidate
fundamentally new information unavailable through traditional FT analysis. While our methods are
generally applicable to a wide range of microscopy techniques, in this work we focus on its
application to STM.

\begin{figure*}[ht!]
	\includegraphics[width=0.98\textwidth]{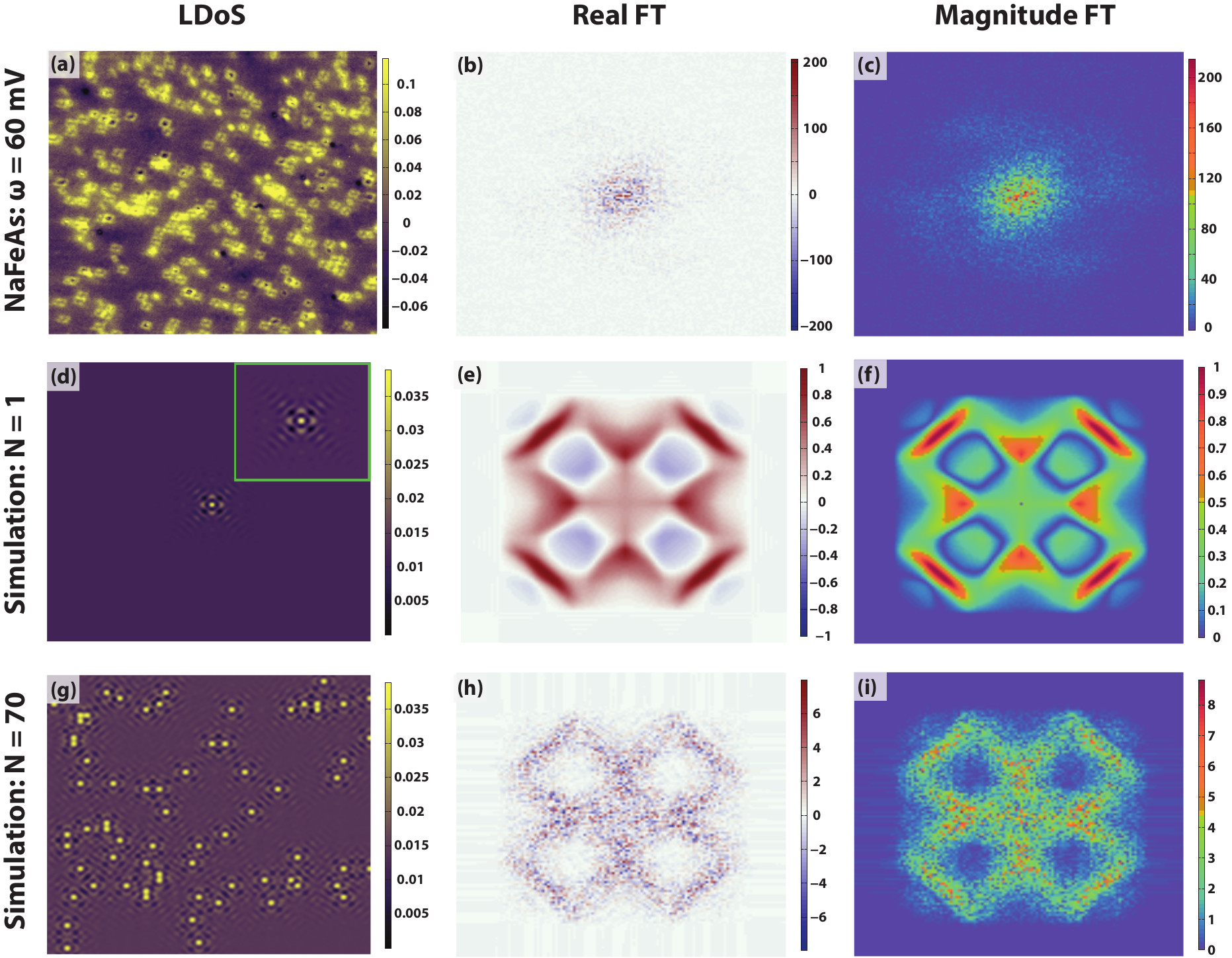}
	\linespread{1}\selectfont{}
	\caption{ 
		Shortcomings of the Fourier transform analysis. 
		(a) The experimentally-obtained LDoS of \NFA{} over a $100 \times 100$ nm$^2$ area of the sample
		at energy $\omega = 60$ mV from Ref~\onlinecite{Rosenthal2014} with junction conditions $V =
		-100$ mV and $I = 300$ pA measured at $T = 26$ K. (b)-(c) Real part and magnitude, respectively,
		of the FT of (a) showing large phase noise. (d) Simulated image of scattering from a single point
		defect. (e)-(f) Real part and magnitude, respectively, of the FT of (d). The LDoS scale in (d)
		was chosen such that the FT magnitude for the simulated single defect had a maximum of 1. (g)-(i)
		Corresponding simulated LDoS and FTs for scattering from 70 point defects that are randomly
		distributed in space. All FT spectra are shown for $-3\pi/5a \le k_x,k_y \le 3\pi/5a$. Comparison
		with the single defect case shows that noise in the FTs arises from the random placement of
		defects. Furthermore, the maximum amplitude of the FT scales approximately as $\sqrt{N}$,
		resulting in a loss of signal fidelity in the FT analysis.
	}
	\label{fig:FTComparison}
\end{figure*}

STM spectroscopy produces two-dimensional spectroscopic maps of the Local Density of States (LDoS)
at position $\vec{x}$ with energy $\omega$, forming a three-dimensional dataset. The contrast in
these images stems from local spatial variations of the LDoS, denoted as $\delta
\rho(\vec{x},\omega)$.  Measurements in which $\delta \rho(\vec{x}, \omega)$ is ascribed to material
defects that cause electron scattering and interference~\cite{Crommie1993} are particularly
interesting, and such maps are often called Quasiparticle Interference (QPI) maps. Analysis of QPI
maps has uncovered information on the dispersion relations and scattering processes in
semimetals~\cite{VonauAubel_PRL_2005, SimonVonau_JPCM_2007}, high-temperature
superconductors~\cite{Hoffman2002, Rosenthal2014, WangLee_PRB_2003, WangLee2011}, and other systems. Let us suppose that the LDoS pattern created by a single defect located at $\vec{x}$ with energy $\omega$
is $\delta \rho_0(\vec{x},\omega)$. This quantity is related to the scattering matrix $\hat{T}$ and bare
Green's function $\hat{G}_0$ through $\delta \rho_0(\vec{x},\omega) = \frac{-1}{\pi} \Im \left[
\bra{\vec{x}}\hat{G}_0\hat{T}\hat{G}_0\ket{\vec{x}} \right]$  (see Supplementary Information Section
I). Accordingly, the STM image from $N$ defects located at $\vec{x}_1,\vec{x}_2, \ldots, \vec{x}_N$
is:
\begin{equation} 
	\label{eq:FT-1} 
	\delta \rho(\vec{x},\omega) = \sum_{j=1}^{N} c_j \delta \rho_0(\vec{x}- \vec{x}_j,\omega)
\end{equation} 
where $c_j$ are constants. A real-world example of such an image is shown in
Figure~\ref{fig:FTComparison}(a), obtained on the pnictide superconductor
\NFA{}~\cite{Rosenthal2014}. 

According to scattering theory, the FT of the QPI image of an
individual defect, $\delta \rho_0(\vec{k},\omega) = \int \diff \vec{x} \,
e^{-i\vec{k}\cdot\vec{x}}\delta \rho_0(\vec{x},\omega)$, is correlated with the underlying
electronic structure of the material~\cite{WangLee_PRB_2003, Heller2003, Kivelson2003}. Most materials of
interest have sufficient disorder so that the LDoS signatures of different defects overlap. In this
situation, it is not possible to identify the isolated defect signature through inspection. Instead,
the traditional analysis~\cite{Arguello2015} proceeds by taking the FT of the entire STM image
$\delta \rho(\vec{x},\omega)$ in~\eqref{eq:FT-1}:
\begin{equation}
	\label{eq:FT-2}
	\delta \rho(\vec{k},\omega) =  \int \diff \vec{x} \, e^{-i\vec{k}\cdot\vec{x}}\delta \rho(\vec{x},\omega) = \delta \rho_0(\vec{k},\omega) \sum_{j=1}^{N} c_j \exp\{-i\vec{k} \cdot \vec{x}_j\}
\end{equation}
While the quantity of interest for QPI analysis is $\delta\rho_0(\vec{k},\omega)$, the experimental
FT image contains a frequency-varying, complex-valued phase factor, $\mathcal{P}(\vec{k}) \equiv
\sum_{j=1}^{N} c_j \exp\{-i\vec{k} \cdot \vec{x}_j\}$. This is illustrated in
Figure~\ref{fig:FTComparison}(b), where the Real Part of the FT (Re-FT) displays wild oscillations
due to $\mathcal{P}(\vec{k})$. To mitigate this, the Magnitude of the FT (mag-FT) is taken, and the
analysis proceeds by assuming that $\mathcal{P}(\vec{k})$ is approximately constant in magnitude so
that $\left| \mathcal{P}(\vec{k})\right| \approx \bar{c} \sqrt{N}$, where $\bar{c}$ is the average
value of the $c_j$. The result of this procedure is illustrated in Figure~\ref{fig:FTComparison}(c),
showing that debilitating noise still persists in the FT after taking the modulus. Moreover, the
procedure of obtaining the mag-FT effectively eliminates half of the useful information in the
complex-valued FT, annihilating all of the phase information from electron scattering processes
originally present in real space. Intense peaks and contours in the real and imaginary parts of
$\delta\rho(\vec{k},\omega)$ are experimental indicators of dominant scattering wavevectors and
order parameter symmetries, which can reveal important properties about the superconducting gap
function sign structure~\cite{Hanaguri_SCGap_Science_2009, ChiJohnston2014_PRB} and surface states
of topological insulators~\cite{ZhangCheng_TopologicalQPI_PRL_2009,
	OkadaDhital_TopologicalQPI_PRL_2011}. However, random phase noise fluctuations in experimental QPI
spectra make comparisons with theoretical QPI calculations
difficult~\cite{HirschfeldAltenfled2015_PRB}.

\begin{figure*}[ht!]
	\includegraphics[width=0.98\textwidth]{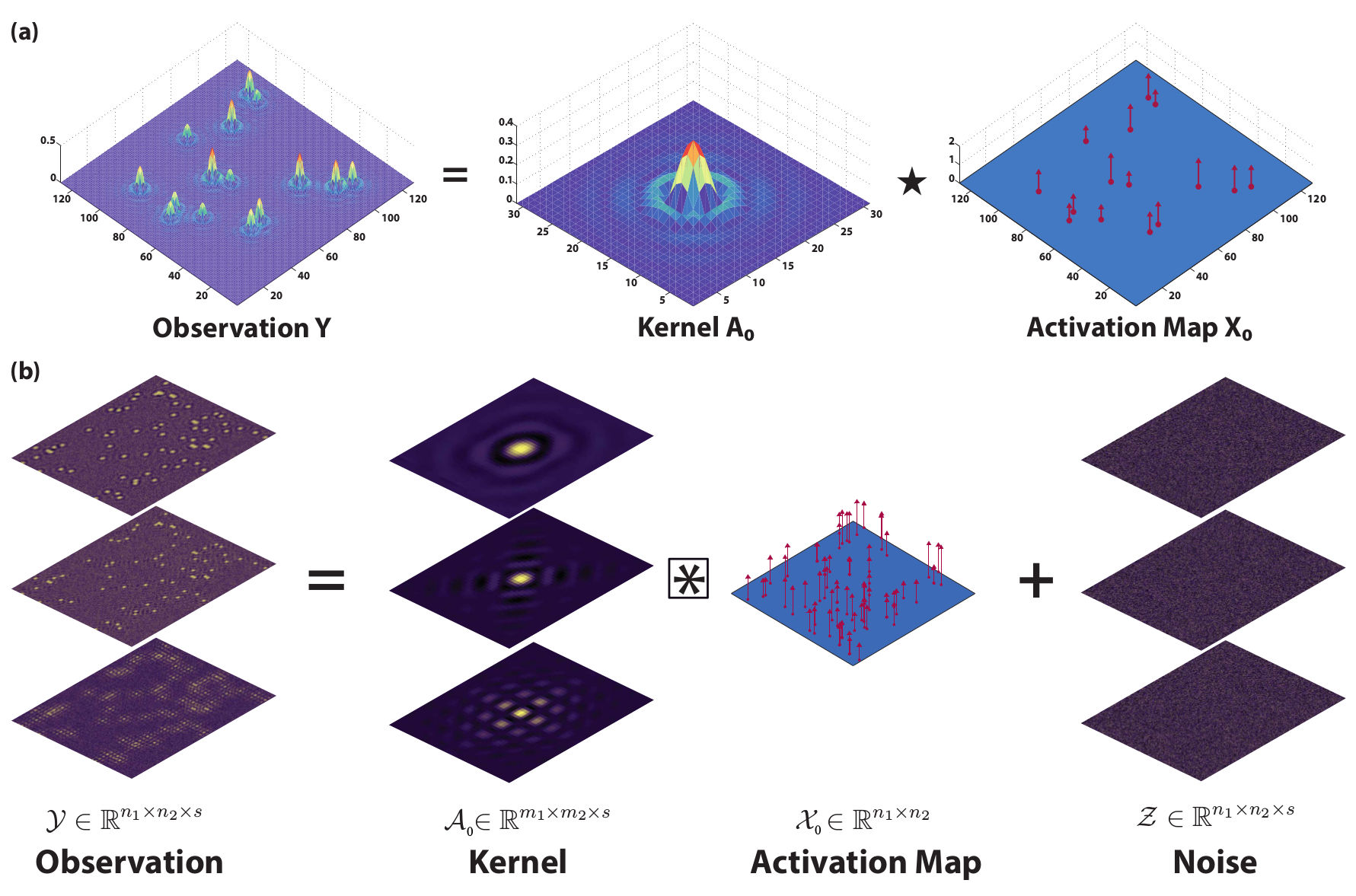}
	\linespread{1}\selectfont{}
	\caption{
		Convolutional data model for STM.
		(a) Schematic representation of the convolutional data model $\mathcal{Y} = \mathcal{A}_0 \star
		\mathcal{X}_0$ for a noise-free simulated STM measurement. (b) Schematic representation of STM
		analysis as a high-dimensional deconvolution problem. Each constant-energy plane of
		$\mathcal{A}_0$ is convolved with the known activation map $\mathcal{X}_0$ and combined
		with additive noise $\mathcal{Z}$ to produce the noisy observation $\mathcal{Y}$. Algorithms for
		SBD seek to efficiently estimate $\mathcal{A}_0$ and $\mathcal{X}_0$ given the noisy measurement
		$\mathcal{Y}$.
	}
	\label{fig:STSDeconvolution}
\end{figure*}

An improved analysis technique to FT-STM would identify the location and the LDoS signature
associated with each defect in a quantitatively-rigorous fashion that respects experimental and
material-specific constraints. For instance, defects remain fixed in position across a series of STM
images in which the measurement bias voltage is varied (see Figure~\ref{fig:STSDeconvolution}). In
this work, we present an analysis technique based on nonconvex optimization that possesses these
desirable features while being broadly applicable to other forms of microscopy and image analysis.

Our algorithm is based on a deconvolutional procedure illustrated in
Figure~\ref{fig:STSDeconvolution}. The image (denoted as $\mathcal{Y}$) in
Figure~\ref{fig:STSDeconvolution}(a) was produced by simulating the effect of quasiparticle
scattering from numerous point defects randomly distributed across the image. At a given bias
voltage, the image consists of a recurrent scattering pattern (called the kernel $\mathcal{A}_0$)
convolved with the locations and relative weights of each defect (called the activation map
$\mathcal{X}_0$) as illustrated in Figure~\ref{fig:STSDeconvolution}(a) and represented as
$\mathcal{Y} = \mathcal{A}_0 \star \mathcal{X}_0$. The underlying challenge in our analysis is to
invert the procedure -- starting with an STM image, determine the kernel and its corresponding
activation map. The kernel and activation map are easily identified by inspection in
Figure~\ref{fig:STSDeconvolution}(a), however this becomes a highly non-trivial problem in the
presence of many overlapping kernels and experimental noise. Similar convolutional models are used
in neuroscience to model neuron spike patterns~\cite{Pillow2013_Neurology} and in systems biology to
capture responses of the endocrine system~\cite{Faghih2014_Biology}. Hence an algorithm developed to
solve this problem has broad applicability.

When $\mathcal{A}_0$ contains multiple slices, with each slice corresponding to a different bias voltage,
we mathematically express the proposed model for STM measurements by collecting the convolutions for
each voltage slice using the notation
\begin{equation}
	\label{eq:LDoSConvolution3}
	\mathcal{Y} = \mathcal{A}_0 \boxast \mathcal{X}_0 + \mathcal{Z},
\end{equation} 
which is schematically depicted in Figure~\ref{fig:STSDeconvolution}(b). The activation map
$\mathcal{X}_0$ is shared globally across all measurement biases and $\mathcal{Z}$ is an additive
noise tensor. The task of recovering both $\mathcal{A}_0$ and $\mathcal{X}_0$ given $\mathcal{Y}$ is
known as the Sparse Blind Deconvolution (SBD) problem.


Over the last decade, a wealth of heuristics and applications for sparse signal recovery have been
developed, often leading to efficient algorithms in theory as well as in
practice~\cite{CandesTao2005_Decoding,Elad2012_SparseRepresentation} (see Supplementary Information
Section II). We investigate the following heuristic for producing estimates $\hat{\mathcal{A}}$ and
$\hat{\mathcal{X}}$\; for $\mathcal{A}_0$ and $\mathcal{X}_0$, by posing an optimization problem
based on \eqref{eq:LDoSConvolution3}:
\begin{equation}
	\label{eq:OptimizationAStar}
	\hat{\mathcal{A}} = 
	\argmin_{\mathcal{A}} 
	\min_{\mathcal{X}} 
	\left[
	\psi_\lambda\left(\mathcal{A}, \mathcal{X} \right) \equiv
	\frac{1}{2} \left\Vert \mathcal{A} \boxast \mathcal{X} - \mathcal{Y} \right\Vert _{F}^{2} + \lambda r ( \mathcal{X} )
	\right]
\end{equation}
which allows one to recover $\mathcal{\hat{X}}=\argmin_{\mathcal{X}} \psi_\lambda(\mathcal{\hat{A}},
\mathcal{X})$.

This is similar to previous formulations proposed for various SBD applications: the Frobenius norm
term $\left\Vert \cdot\right\Vert _{F}^{2}$ promotes data fidelity upon minimization
$(\mathcal{Y}\simeq\mathcal{\hat{A}}\boxast\mathcal{\hat{X}})$, and a regularization term
$r(\hat{\mathcal{X}})$ is chosen, such as the $\ell_1$ norm, so that the minimization encourages
$\mathcal{\hat{X}}$ to be sparse, with $\lambda \geq 0$ governing the trade-off between the two
objectives. However, most SBD applications focus on signal enhancement that uses the convolutional
model as a rough guideline, leading to a weak notion of accurate
estimation~\cite{levin2009understanding}. In contrast, the convolutional model fits naturally into
the STM setting, in which robust, consistent results are paramount for scientific investigation.
These considerations prompt a number of choices that are not emphasized in previous heuristics, such
as the domain of $\mathcal{A}$, form of $r(\hat{\mathcal{X}})$, and refinement of the estimates.

In order to solve this optimization problem, we present the SBD-STM algorithm:

\begin{algorithm}[H]
	\scriptsize
	\begin{flushleft}
		\linespread{1}\selectfont{}
		\caption{Complete SBD-STM Procedure}
		\textbf{Input: }
		\begin{itemize}
			\item Observation \textbf{$\mc{Y}\in\mathbb{R}^{n_{1}\times n_{2}\times s}$}, kernel size $\left(m_{1},m_{2}\right)$, initial $\lambda_{0}\geq0$, decay rate $\alpha\in\left[0,1\right)$,
			and final $\lambda_{\mathrm{end}}\geq0$.
		\end{itemize}
		\textbf{Initial phase:}
		\begin{enumerate}
			\item Randomly initialize: $\mc{A}^{\left(0\right)}\in\mc{S=\mathbb{S}}^{m_{1}\times m_{2}\times s}$.
			\item $\mc{A}_{*}^{\left(0\right)}, \mc{X}_{*}^{\left(0\right)} \leftarrow{\tt ASolve}\left(\mc{A}^{\left(0\right)},\lambda_{0}, \mc{Y}\right)$.
		\end{enumerate}
		\textbf{Refinement phase:}
		\begin{enumerate}
			\item Lifting: Get $\mc{A}^{\left(1\right)}\in S^{'}=\mc{\mathbb{S}}^{m_{1}^{'}\times m_{2}^{'}\times s}$
			by zero-padding the edges of $\mc{A}_{*}^{\left(0\right)}$ with
			a border of width $\left\lfloor \frac{m_{i}}{2}\right\rfloor$.
			\item Set $\lambda_{1}=\lambda_{0}$.
			\item Continuation: \textbf{Repeat} for $k=1,2,\dots$ \textbf{until} $\lambda_{k}\leq\lambda_{\mathrm{end}}$,
			\begin{enumerate}
				\item $\mc{A}_{*}^{\left(k\right)}, \mc{X}_{*}^{\left(k\right)} \leftarrow{\tt ASolve}\left(\mc{A}^{\left(k\right)},\lambda_{k}, \mc{Y}, \mc{X}_{*}^{\left(k-1\right)}\right)$,
				\item Centering: 
				\begin{enumerate}
					\item Find the size $m_{1}\times m_{2}$ submatrix of $\mc{A}_{*}^{\left(k\right)}$ that maximizes the Frobenius (square) norm across all $m_{1}\times m_{2}$ submatrices.
					\item Get $\mc{A}^{\left(k+1\right)}$ by shifting $\mc{A}_{*}^{\left(k\right)}$
					so that the chosen $m_{1}\times m_{2}$ restriction is in the center,
					removing and zeropadding entries as needed.
					\item Normalize $\mc{A}^{\left(k+1\right)}$ so it lies in $\mc{S}^{'}$.\
					\item Shift $\mc{X}_{*}^{\left(k\right)}$ along the anti-parallel vector to the shift of $\mc{A}_{*}^{\left(k\right)}$.
				\end{enumerate}
				\item Set $\lambda_{k+1}=\alpha\lambda_{k}$.
			\end{enumerate}
		\end{enumerate}
		\textbf{Output:}
		\begin{enumerate}
			\item Extract $\hat{\mc{A}}\in\mc{S}$ by extracting the restriction
			of the final $\mc{A}^{\left(k+1\right)}$ to the center $m_{1}\times m_{2}$
			window.
			\item Find the corresponding activation map $\hat{\mc{X}}\in\mathbb{R}^{n_{1}\times n_{2}}$
			by solving $\min_{\mc{X}}\psi_{\lambda_{k}}\big(\hat{\mc{A}},\mc{X}\big)$.
		\end{enumerate}
		\Algphase{Function \tt Asolve}
		\textbf{Input:}
		\begin{itemize}
			\item Current kernel, $\mc{A}_{in}$, current sparsity parameter, $\lambda_{in}$, the observation $\mc{Y}$, current activation map $\mc{X}_{in}$ (Refinement Phase)
		\end{itemize}
		\textbf{Minimization}
		\begin{algorithmic}
			\If {Initial Phase}
			\State $\mc{X}_1 \leftarrow$ Minimize $\mc{X}$ for $\psi_{\lambda_{in}}(\mc{A}_{in}, \mc{X})$ using the FISTA algorithm \cite{FISTA}.
			\Else
			\State $\mc{X}_1 \leftarrow \mc{X}_{in}$
			\EndIf
		\end{algorithmic}
		\begin{enumerate}
			\item $\mc{A}_{out} \leftarrow$ Minimize $\mc{A}$ for $\psi_{\lambda_{in}}(\mc{A}, \mc{X}_{1})$ using the Riemannian Trust-Region Method (RTRM) over the sphere \cite{RTRM}.
			\item $\mc{X}_{out} \leftarrow$ Minimize $\mc{X}$ for $\psi_{\lambda_{in}}(\mc{A}_{out}, \mc{X})$ using FISTA.
		\end{enumerate}
		\textbf{Output:}
		\begin{itemize}
			\item $\mc{A}_{out}$, $\mc{X}_{out}$.
		\end{itemize}
	\end{flushleft}
\end{algorithm}

See Supplementary Information Sections II and III for further discussion on formulating and
solving~\eqref{eq:OptimizationAStar}, and Section IV for how our approach to SBD can be applied to
image deblurring by using an objective similar to \eqref{eq:OptimizationAStar}.

\begin{figure*}[ht!]
	\includegraphics[width=.98\textwidth]{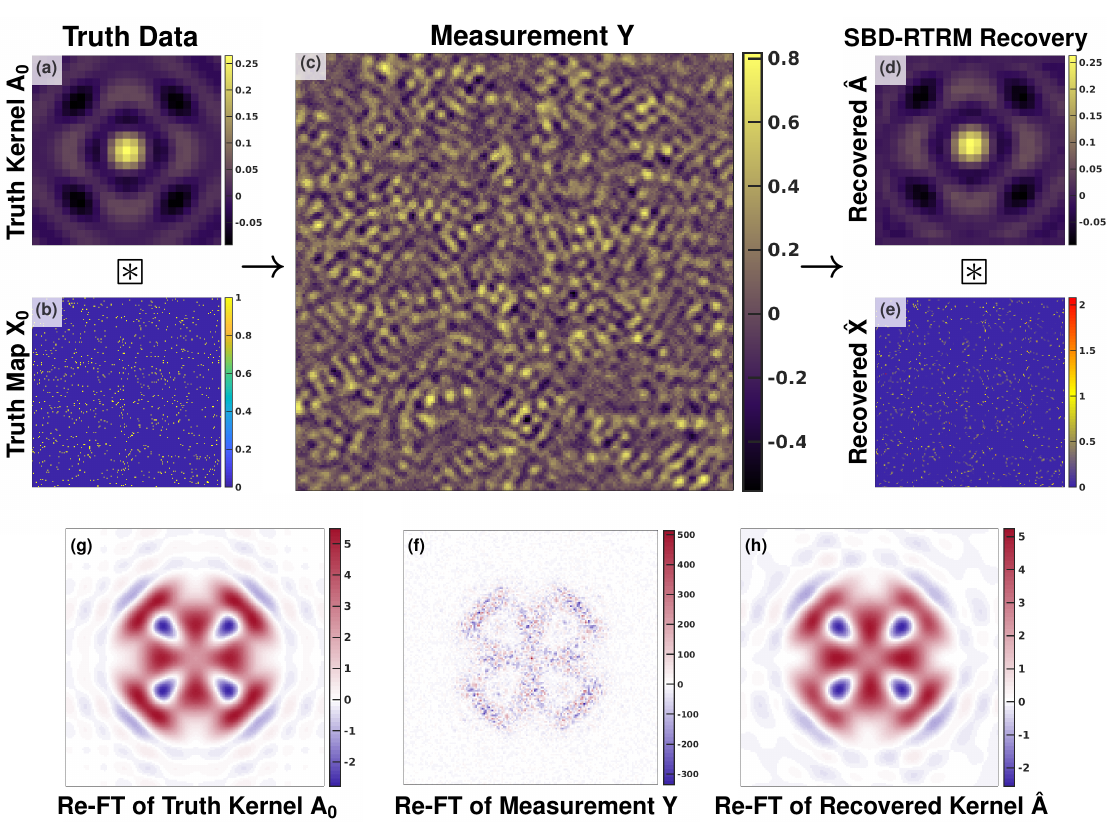}
	\linespread{1}\selectfont{}
	\caption{
		Illustrating the SBD-STM procedure.
		The simulated kernel (size $25 \times 25$ pixels) in (a) is convolved with the activation map in
		(b) and added to random noise with $\textrm{SNR} \approx 0.792$ to produce the simulated noisy
		measurement shown in (c). The simulated measurement is a $185 \times 185$ pixel image with a
		defect concentration of $\theta = 2.73\%$. (d) Recovered kernel and (e) recovered activation map
		obtained using SBD-STM with $\lambda = 0.1$, showing excellent agreement with simulation inputs in
		(a) and (b). The Re-FTs of the truth kernel, noisy measurement, and the recovered kernel from
		SBD-STM, shown in panels (f), (g), and (h), respectively. The Re-FT spectra are all shown with
		$-3\pi/5a \le k_x,k_y \le 3\pi/5a$. The SBD-STM based FT shows both vastly improved SNR and a
		phase-sensitive recovery of the scattering interference.
	}
	\label{fig:STS-OneExampleSim}
\end{figure*}

To demonstrate the strength of SBD-STM, consider the situation illustrated in
Figure~\ref{fig:STS-OneExampleSim}. We generated a simulated observation $\mathcal{Y}$ using a
ground truth scattering pattern $\mathcal{A}_0$ similar to Figure~\ref{fig:FTComparison}(d) and a
dense, randomly generated activation map $\mathcal{X}_0$ shown in
Figure~\ref{fig:STS-OneExampleSim}(b). Convolving the truth data with the activation map and adding
significant white noise with variance $\eta$ so that the Signal-to-Noise Ratio (SNR) is less than
unity, with $\textrm{SNR} \equiv \frac{var(\mathcal{A}_0)}{\eta}$, we generate the image
shown in Figure~\ref{fig:STS-OneExampleSim}(c). With many overlapping kernels and substantial noise,
it is a futile task to accurately identify the underlying kernel and activation map of the image
through visual inspection. However, as shown in Figure~\ref{fig:STS-OneExampleSim}(d)-(e), SBD-STM
successfully recovers a kernel $\hat{\mathcal{A}}$ and its associated activation map
$\hat{\mathcal{X}}$ that closely resemble the truth data. The results shown in
Figure~\ref{fig:STS-OneExampleSim} were obtained with a fixed $\lambda = 0.1$. The scaling of the activation map entries is due to the choice of $\lambda$, and the noise also introduces blurring in the activation map \cite{Reweighting}. Despite this, the overall features of the activation map and the recovered kernel are remarkably similar to the ground truth.

 In Fourier space, the Re-FT of $\mathcal{Y}$ is missing crucial features
of the true Re-FT spectrum and has noise fluctuations $\approx 100$ times that of the true transform.
However, the Re-FT of the SBD-STM recovered kernel is consistent with the true Re-FT in both its
structure and amplitude.

In our implementation, SBD-STM yields an activation map shared across all bias voltages. This not
only reveals the spatial distribution of the defect kernels, but also naturally improves the
accuracy of the SBD-STM recovered kernels at bias energies with noisy measurements. Consequently,
SBD-STM returns more physically meaningful results when data from multiple biases are simultaneously
analyzed than if each constant-bias slice of $\mathcal{Y}$ were individually analyzed. SBD-STM
results on a simulated noisy STM dataset with 41 bias voltages are found in Supplementary
Information Section V, demonstrating that SBD-STM is successful in optimizing the objective function
with STM constraints in mind.

\begin{figure*}[ht!]
	\centering
	\includegraphics[width=0.95\textwidth]{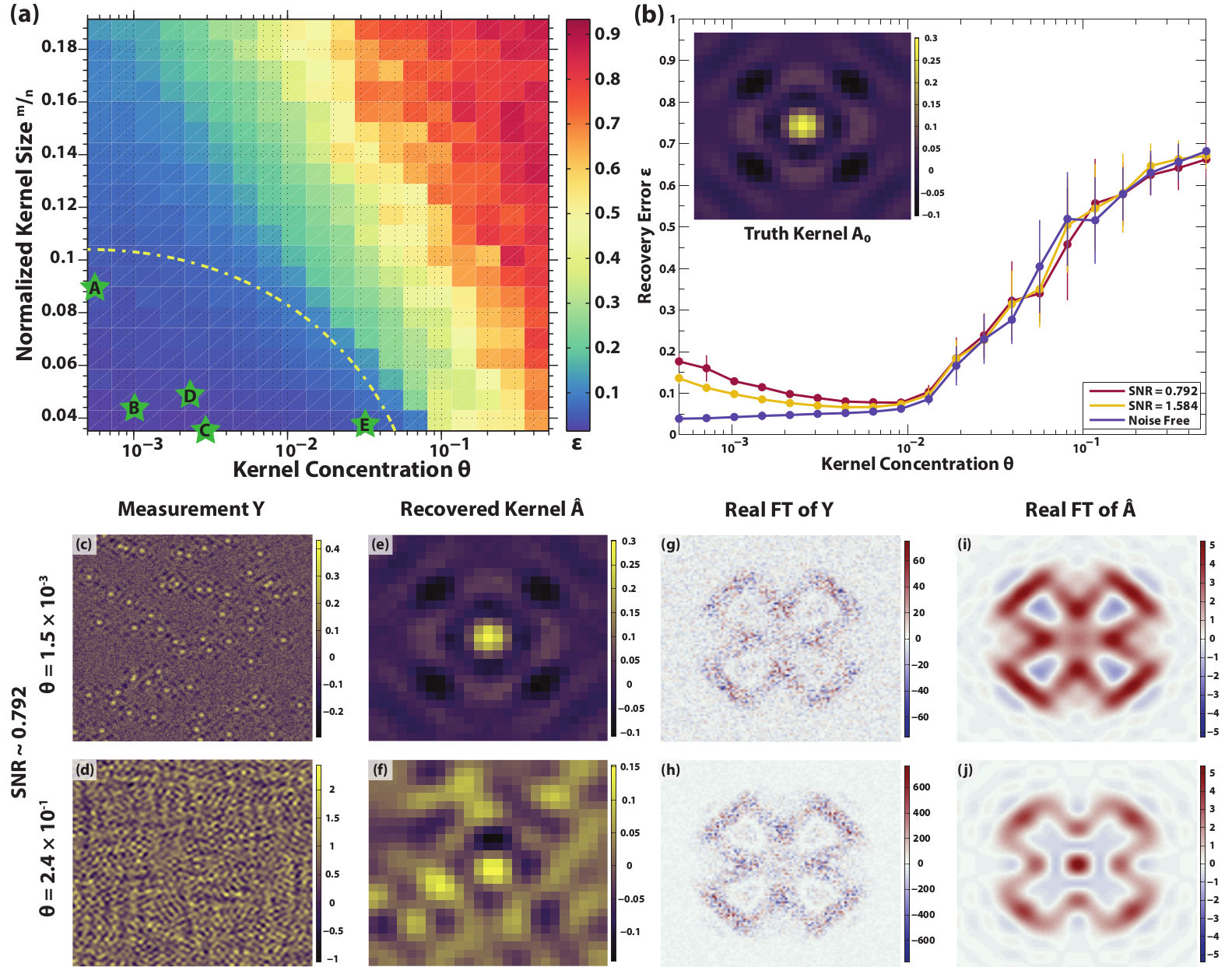}
	\linespread{1}\selectfont{}
	\caption{ 
		Benchmarking SBD-STM performance. 
		(a) SBD-STM performance phase diagram from noise-free simulated results on images of area $n = 256
		\times 256$, plotting the error metric $\epsilon(\hat{\mathcal{A}}_{\theta,\eta},\mathcal{A}_0 )$
		as a function of normalized kernel size and concentration $\theta$ (the number of defects scaled
		by the number of pixels). Each point on the diagram is determined from the mean of 20 independent
		simulated measurements. SBD-STM performs well in the blue regions, but begins to fail in the red
		regions, at very high kernel size or concentration. The region where typical STM measurements are
		performed is bounded by the dashed line and the lower and left axes. Green stars labeled A, B, C,
		D, and E indicate representative images from Refs.~\onlinecite{Stroscio2007, RoushanSeo_Nature_2009,
			Rosenthal2014, Arguello2015, AllanChuang2013}, respectively. (b) Performance of SBD-STM in the
		presence of additive noise in the measurement at a fixed value of the normalized kernel size
		$m/n=0.14$.  (c)-(f) Examples of simulated measurements from panel (b) and their corresponding
		SBD-STM recovered kernels. (g)-(j) A comparison of the Re-FTs of the measurement and the SBD-STM
		recovered kernel, indicating that SBD-STM consistently outperforms FT-STM across all parameter
		ranges. All Re-FT spectra are shown with $-3\pi/5a \le k_x,k_y \le 3\pi/5a$.
	}
	\label{fig:SimulationSummary1}
\end{figure*}

Before invoking SBD-STM on experimental data, we must understand its limitations and domain of
applicability. The complexity of the STM deconvolution problem varies depending on the SNR and the
overlap tendency of nearby defects. A series of numerical experiments on simulated data was
performed to investigate the effects of defect concentration $\theta$ -- the probability that any
entry of $\mathcal X_0$ is ``on'' -- and additive measurement noise on the expected success of
SBD-STM. Simulated STM images are produced in a similar fashion as in
Figure~\ref{fig:STS-OneExampleSim}, and the performance of SBD-STM was assessed as a function of
four adjustable parameters - the image size $n \equiv n_1\times n_2$, kernel size $m \equiv m_1
\times m_2$, kernel concentration $\theta$, and SNR. Details of the data generation and simulation
work are contained in Supplementary Information Sections I and VI. To assess the accuracy of kernel
recovery in real space, we define the real space recovery error metric as
$\epsilon(\hat{\mathcal{A}}_{\theta,\eta},\mathcal{A}_0 ) \equiv \frac{2}{\pi} \arccos \left|
\langle \hat{\mathcal{A}}_{\theta,\eta}, \mathcal{A}_0 \rangle \right|$, with
$\langle\hat{\mathcal{A}}_{\theta,\eta}, \mathcal{A}_0\rangle$ denoting the inner product of the
vectorizations of $\hat{\mathcal{A}}_{\theta,\eta}$ and $\mathcal{A}_0$, which are the recovered and
truth kernels, respectively. Figure~\ref{fig:SimulationSummary1}(a) depicts a normalized defect size
$\frac{m}{n}$ vs. concentration $\theta$ ``phase diagram'' to explore the interplay between
$\frac{m}{n}$ and $\theta$ on real-space algorithmic accuracy
$\epsilon(\hat{\mathcal{A}}_\theta,\mathcal{A}_0 )$ in noise-free $(\eta = 0)$ simulated
measurements. We observe a phase transition in SBD-STM performance in the $\frac{m}{n}-\theta$
plane. The bottom-left of the plot captures situations in which the defects have negligible
probability of overlapping, facilitating the near-perfect deconvolution of the noise free image.
Increasing either  $\frac{m}{n}$ or $\theta$ introduces error in the kernel recovery due to
increased overlapping between defects. Practically, $\frac{m}{n}$ can be reduced by increasing the
overall STM measurement area $n$ in an attempt to perform deconvolution-by-inspection. However at
high defect concentrations $\theta$ or moderate noise levels, this strategy cannot guarantee success
while SBD-STM can still return reliable estimates.

Next, we briefly discuss the SBD-STM performance as a function of noise in the signal.
Figure~\ref{fig:SimulationSummary1}(b) shows the evolution of
$\epsilon(\hat{\mathcal{A}}_{\theta,\eta},\mathcal{A}_0 )$ as a function of defect concentration
$\theta$ for three values of SNR ranging from noise free to noise-dominated (SNR $ = 0.792$). At
high noise levels, performance error fluctuations in the far left of the plot appear because of the
statistically futile challenge of accurately identifying low-density, low-intensity motifs under
high levels of noise. The error curves appear to converge into a narrow band when $\theta \gtrsim
0.01$, demonstrating that the algorithm is robust to a wide range of SNRs for higher concentrations.
By $\theta \approx 0.2$, the defect concentrations are sufficiently dense so that virtually all
defect kernels are overlapping, causing SBD-STM to collapse and return unreliable estimates. These
trends persist when the kernel size and SNR are further increased, as described in Supplementary
Information Section VI. Altogether, our simulations show that in the range of parameters where
typical STM experiments are performed (bounded by the dashed line in
Figure~\ref{fig:SimulationSummary1}(a)), SBD-STM performs splendidly ($\epsilon < 0.1$) and
consistently outperforms the usual FT-STM methodology even in the presence of considerable noise, as
seen in Figure~\ref{fig:SimulationSummary1}(c)-(j).

\begin{figure*}[ht!]
	\centering
	\includegraphics[width=0.66\textwidth]{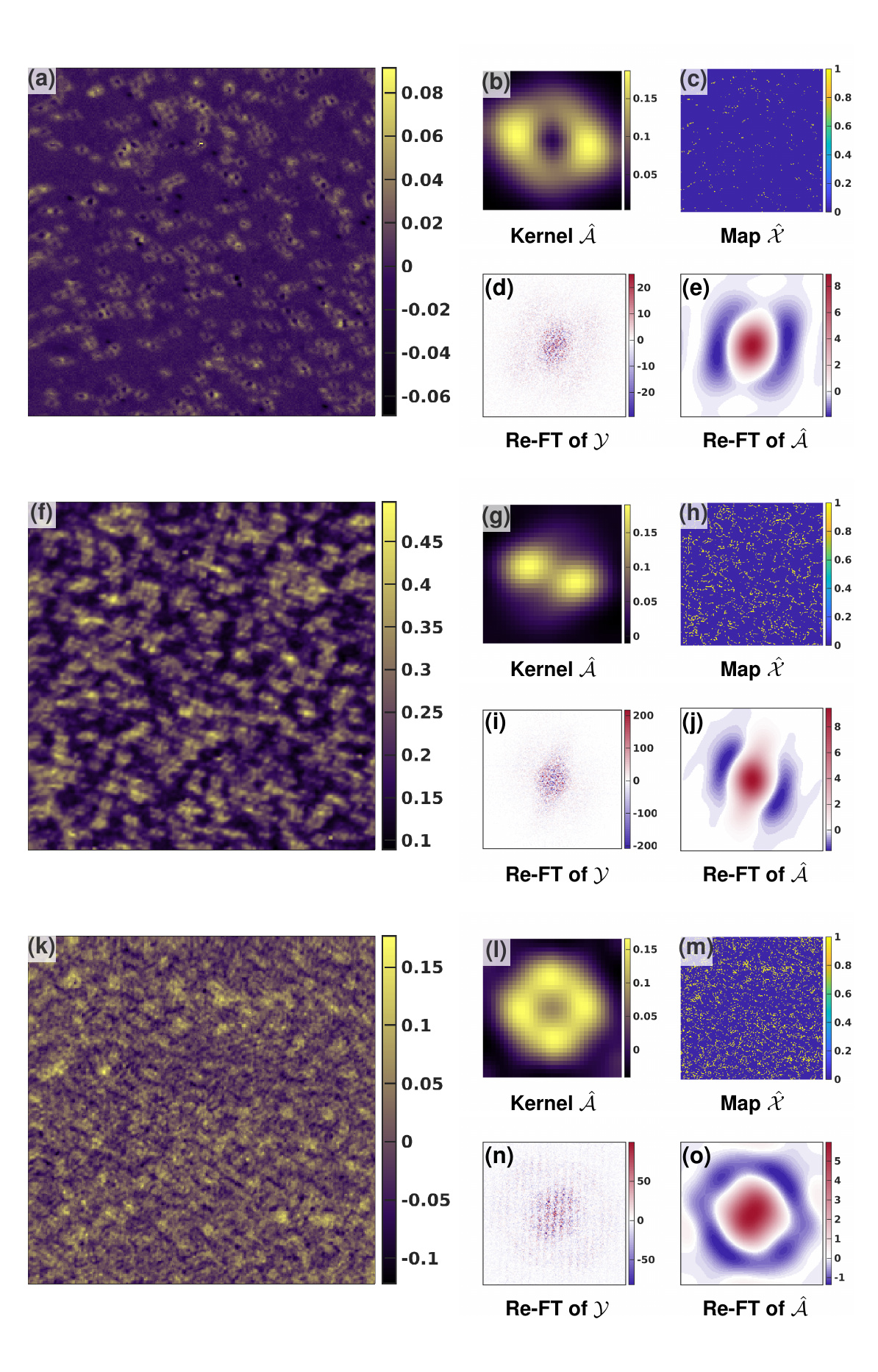}
	\linespread{1}\selectfont{}
	\caption{SBD-STM recoveries with increasing density of defects. In the left column, we have conductance maps with $\mathrm{Co}$ defects in \NFCA{} of (a) high-quality ($T = 71$ K, $I = 59$ pA, $V_{set} = -100$ mV, $114 \times 114$ nm), (f) $x = 0.015$ ($T = 6$ K, $I = -200$ pA, $V_{set} = -20$ mV, $50.4 \times 50.4$ nm), and (k) $x = 0.02$ ($T = 5.9$ K, $I = 200$ pA, $V_{set} = -50$ mV, $250 \times 250$ nm) biased at energies of (a) $40$ meV, (f) $5$ meV, and (k) $-7$ meV. In the right column, clockwise from the top left, we have the recovered kernel (b,g,l), the recovered activation map (c,h,m), the Re-FT of the kernel (d,i,n), and the Re-FT of the conductance map (e,j,o), for comparison. The recovered kernels have been given 2-fold rotational symmetry, and all activations have been set to 1 to make them more readily visible.}
	\label{fig:NaFeAs-Results}
\end{figure*}

\textcolor{black}{The results obtained above on synthetic data show that SBD-STM is able to recover kernels even at high defect density and in the presence of significant white noise, where alternate techniques such as manual detection fail. However, one might still question whether the method will work on real STM data, which might have other types of noise or errors which we have not accounted for in our synthetic data. In order to investigate this fully, we now apply SBD-STM to investigate a set of experimentally-obtained STM images of
\NFCA{} at different values of $x = 0.0$, $0.015$ and $0.02$. At $x = 0.0$ (parent compound), no additional cobalt dopants are present in the lattice, and the only defects present are those intrinsic to the crystal, which are at a concentration of about 1\%. This compound has been previously studied via STM and the data have been analyzed using the standard Fourier transform technique ~\cite{Rosenthal2014}. To contrast with the standard FT-STM analysis, we implement SBD-STM on raw experimental data to demonstrate the significant improvement in data fidelity of the Re-FT. Shown in figure ~\ref{fig:NaFeAs-Results} (b) and (c) are the recovered Kernel and activation map respectively from the map shown in (a). Figure ~\ref{fig:NaFeAs-Results} (d) and (e) show the Re-FT of the entire image and of the kernel respectively over the same range in Fourier space. The phase sensitive recovery of the Fourier transform in Figure~\ref{fig:NaFeAs-Results} (e) when compared to (d) is immediately apparent. We note that for this particular doping the individual kernels are well separated, and the kernel can be isolated directly by eye from the large area image in Figure~\ref{fig:NaFeAs-Results}(a). We show the application of SBD-STM to this sample to illustrate that the recovered kernel is indeed what one would expect by direct measurement around an individual defect.} 

\textcolor{black}{We now turn to the sample with $x = 0.015$, a STS image of which is shown in Figure~\ref{fig:NaFeAs-Results}(f). At this doping, we see that there are regions of high-density clustering of the kernels, where the clustering is dense enough such that the individual kernels overlap and become difficult to resolve. In Figure~\ref{fig:NaFeAs-Results}(g) and (h), we show the corresponding recovery for the kernel and activation map respectively. The Re-FT of the entire map and the recovered kernel are shown in Figure~\ref{fig:NaFeAs-Results}(i) and (j) respectively. At this defect level, we can very occasionally see isolated defects, and the recovered kernel is seen to nicely match with the differential conductance around these isolated defects. }  

\textcolor{black}{Finally,  we consider a sample that is optimally doped (highest $T_c$) with a nominal cobalt concentration of $x = 0.02$. Shown in Figure~\ref{fig:NaFeAs-Results}(k) is an STS image obtained on this sample at $T = 5.9$ K. At this doping level, there are no isolated impurities present anywhere in the sample, and the kernel can not be manually recovered. The SBD-STM is able to recover the kernel and activation map as shown in  Figure~\ref{fig:NaFeAs-Results}(l) and (m) respectively. The Re-FT of the entire image and of the recovered kernel are shown in Figure~\ref{fig:NaFeAs-Results}(n) and (o) respectively. We see from the series of data in Figure~\ref{fig:NaFeAs-Results} that SBD-STM works over the entire doping range that is relevant to STM experiments and is able to recover high-quality kernels with phase-sensitive Fourier transforms. As a side benefit of SBD-STM, we can obtain more accurate defect or dopant counts from the recovered activation maps.}

\textcolor{black}{From the results shown on both synthetic and real experimental STM data, we can see that SBD-STM provides a complete recovery of kernels in real space and therefore a phase sensitive recovery of the Fourier transform in reciprocal space. Within the formulation of quasiparticle interference, the phase of the Fourier transform in the QPI signal is dependent on the incoming and outgoing quasiparticle's Green's function as well as the potential of the impurity. The availability of phase information can give us new insight into individual materials that is not available simply from the magnitude of the QPI signal. In the remainder of this paper, we consider one such new insight into the physics of \NFCA{}. \NFCA{}, like many of the pnictides displays a superconducting dome as a function of cobalt doping. The maximum $T_c$ at optimal doping ($x = 0.02$) reaches 18 K. As with the other pnictides, determining the symmetry of the superconducting order parameter in this compound is of much current interest.  In this context, recent theoretical work on QPI in the superconducting state of the pnictides \cite{Hirschfeld, Hirschfeld2} has opened up the possibility of distinguishing different superconducting order parameters from their QPI signature under the assumption that interband scattering in the superconducting state dominates the QPI signal.}
\textcolor{black}{In \cite{Hirschfeld}, the (real) Fourier transform of the QPI signal around a single defect $ \delta\rho(\vec{q},\omega)$ is integrated over momentum to produce the quantity $ \delta\rho(\omega)$. This quantity is then anti-symmetrized with respect to energy relative to the Fermi level to produce a quantity $ \delta\rho^{-}(\omega) = \delta\rho(\omega) - \delta\rho(-\omega)$. It is shown in  \cite{Hirschfeld} that in the case of $s^{\pm}$ pairing, $ \delta\rho^{-}(\omega)$ is large and of constant sign over the energy range near the superconducting gap. Conversely, in the case of $s^{++}$ pairing, $\delta\rho^{-}(\omega)$ is expected to be small and have a sign change around the gap energy. In order to carry out the integration over $\vec{q}$ described in this procedure, it is required that we have access to the phase of the QPI signal.} \textcolor{black}{One way of achieving this is to directly image around an isolated dopant or defect  where the complete phase sensitive pattern can be measured. Such STS imaging has recently been performed on iron chalcogenides \cite{Sprau75, Du2017}. However, this method has not been applied to the iron arsenides, especially at optimal doping where the defect or dopant density is high and the phase information in the FT was not previously available. Armed with the phase-sensitivity of SBD-STM, we now analyze differential conductance maps of near optimally doped \NFCA to investigate the superconducting order parameter at optimal doping. }

\begin{figure}
	\centering
	\includegraphics[width=1\textwidth]{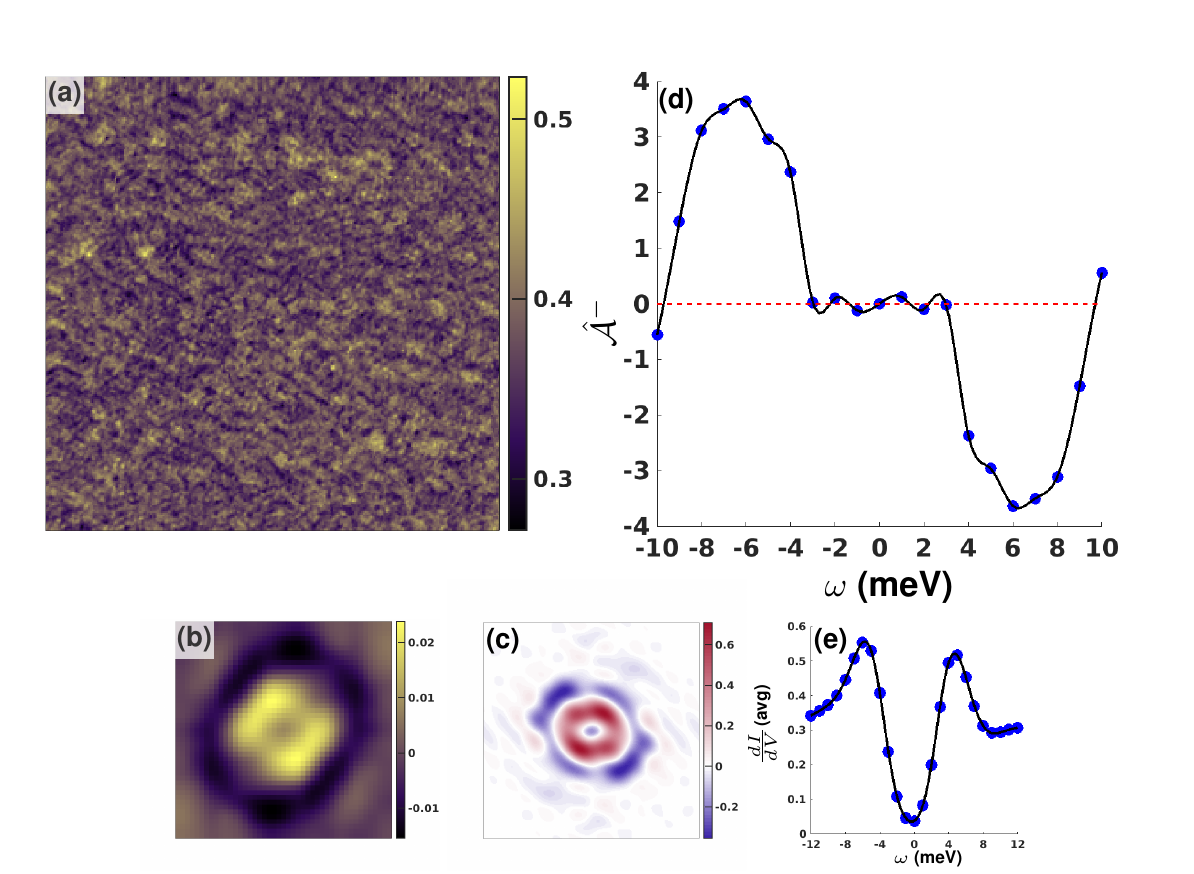}
	\caption{ (a) Differential conductance map at $\omega=-10$ meV for an optimally doped \NFCA sample (x=0.02), over a 250 nm $\times$ 250 nm field of view with $T = 5.9$ K, $I = 200 pA$, and $V_{set} = -50$ meV. b) recovered kernel, $\mathcal{\hat{A}}$, from the conductance map in a). c) Fourier transform of the recovered kernel.  d) $\mathcal{\hat{A}}^{-}(\omega)$, computed from the recovered kernels. The large and constant sign of the response is an indication of $s^{\pm}$ pairing.  e) Spatially averaged differential conductance from the same area as a function of energy, revealing the superconducting gap 2$\Delta=11$   meV.}
	\label{Hirschfeld}
\end{figure}

We start with a dataset that consists of 21 raw STS images from -10 meV to +10 meV in 1 meV increments on an optimally doped \NFCA sample at T=5.9 K. One of these raw images at $\omega=-10$ meV is shown in Figure \ref{Hirschfeld}(a). Notice that no individual motifs can be resolved by eye. We then proceed to recover the kernels $\mathcal{\hat{A}}(\vec{r},\omega)$ at each energy using SBD-STM.  An example of this recovery is shown in Figure \ref{Hirschfeld}(b), which is the recovered SBD-STM kernel $\mathcal{\hat{A}}(r,\omega=-10$ meV), recovered from the raw STS image in Figure \ref{Hirschfeld}(a). The recovered kernel lacks the strong anisotropy of recovered kernels from the underdoped regime, suggesting that electronic nematicity is not strong at this doping. Assuming that SBD-STM has worked correctly, the recovered $\mathcal{\hat{A}}(\vec{r},\omega)$ is identical to the real space QPI signal $ \delta\rho(\vec{r},\omega)$.  We then take the real-part of the 2-D Fourier transform $\mathcal{\hat{A}}(\vec{q},\omega)$, as shown in Figure \ref{Hirschfeld}(c) at $\omega=-10$ meV. This FT has the full phase information present, and we can then integrate over $\vec{q}$ and antisymmetrize with respect to energy:

$$  \mathcal{\hat{A}^{-}}(\omega) = \sum_{\vec{q}} Re(\mathcal{\hat{A}}(\vec{q},\omega)) - Re(\mathcal{\hat{A}}(\vec{q}, -\omega))$$

We perform this procedure at each energy, and plot the resultant $\mathcal{\hat{A}}^{-}(\omega)$ in Figure \ref{Hirschfeld}(d). In (e), we show the spatially averaged differential conductance from the same data sets as a function of energy, revealing the superconducting gap. From the two coherence peaks, we calculate a $2\Delta$ of 11 meV. We can clearly see from Figure \ref{Hirschfeld}(d) that $\mathcal{\hat{A}}^{-}(\omega)$ is peaked near the superconducting gap, and has no sign change in the energy range near the gap, as expected for an $s^{\pm}$ order parameter. This procedure illustrates some of the new physical insight into STS image data that can be obtained once the complete phase information in the QPI signal is available for analysis.

In its current implementation, SBD-STM addresses the problem of identifying a single motif across a
series of images. Beyond the identification of real space motifs in microscopy images, SBD-STM can
also be applied to problems in which the motif is sparse in the appropriately chosen space, such as
sparsity in the spatial gradient for natural image deblurring~\cite{Fergus2006-ACM,Levin2011-PAMI}.
Moreover, the flexibility of the convolutional data model in~\eqref{eq:LDoSConvolution3} affords the
natural generalization, $\mathcal{Y} = \sum_{j=1}^M \mathcal{A}_0^{\left(j\right)} \boxast
\mathcal{X}_0^{\left(j\right)} + \mathcal{Z}$, which expands the scope of SBD-STM to identify
multiple distinct kernels in any series of images. In particular, STM images that contain various
short-range orders, such as charge or spin density waves, would be amenable to a similar
analysis~\cite{Fujita2014_PhaseSensitive, Arguello2015, AllanChuang2013, CaiZhou2013_NaFeCoAs,
	HamidianEdkins2016_dffDW}. SBD-STM recovered results from these STM measurements can be directly
compared with theoretical predictions~\cite{KnolleEremin2010_PRL,
	WangAgterberg2015_PRL_CompetingOrderCuprate, SchattnerGerlach2016_PRL_CompetingOrderAFM} to
understand the nature of competing orders in superconductors and other strongly correlated
materials. Other analysis methodologies~\cite{Fujita2014_PhaseSensitive,
	DallaTorreHe2016_Holographic, HamidianEdkins2016_dffDW} have been recently proposed to improve FT
data fidelity and provide some phase-sensitive information on the structure of ordered phases. These
alternative approaches provide compelling information under suitable conditions, but their results
are still vulnerable to phase noise contamination. The correct implementation of SBD-STM to such
cases remains an open but solvable problem.

We thank Ethan Rosenthal and Erick Andrade for help with STM data acquisition, and Andrew Millis and
Rafael Fernandes for discussions. This work is supported by the National Science Foundation Bigdata
program (Grant number IIS-1546411). Support for STM equipment and operations is provided by the Air
Force Office of Scientific Research (Grant number FA9550-16-1-0601).

\bibliography{STM_BD_NatComm_Rev_v2}

\end{document}


\title{Dictionary Learning in Fourier Transform Scanning Tunneling Spectroscopy - 
	Supplementary Information}

\author{Sky C. Cheung \footnote{These authors contributed equally to this work}}
\affiliation{Department of Physics, Columbia University, New York, NY 10027 USA}
\author{John Y. Shin \footnotemark[1]}
\affiliation{Department of Physics, Columbia University, New York, NY 10027 USA}
\author{Yenson Lau}
\affiliation{Department of Electrical Engineering, Columbia University, New York, NY 10027 USA}
\author{Zhengyu Chen}
\affiliation{Department of Electrical Engineering, Columbia University, New York, NY 10027 USA}
\author{Ju Sun}
\affiliation{Department of Electrical Engineering, Columbia University, New York, NY 10027 USA}
\author{Yuqian Zhang}
\affiliation{Department of Electrical Engineering, Columbia University, New York, NY 10027 USA}
\author{John N. Wright}
\email{johnwright@ee.columbia.edu}
\affiliation{Department of Electrical Engineering, Columbia University, New York, NY 10027 USA}
\author{Abhay N. Pasupathy}
\email{pasupathy@phys.columbia.edu}
\affiliation{Department of Physics, Columbia University, New York, NY 10027 USA}

\date{\today}



\maketitle

\newpage

\begin{table}[ht!]
	\begin{ruledtabular}
		\centering
		\linespread{1}\selectfont{}
		\caption{Summary of symbols and notation}
		\begin{tabular}{ p{2.5cm} l } 
			Symbol & Description \\ \hline
			$\vec{x}$ & position: $\bbR^2$ \\
			$\vec{k}$ & wavevector (Fourier dual to $\vec{x}$): $\bbR^2$ \\
			$\omega$ & energy level (applied bias voltage): $\bbR$ \\
			$\rho(\vec{x},\omega)$ & LDoS/QPI map: $\bbR^2\times\bbR\rightarrow\bbR$ \\
			$\rho_0(\vec{x},\omega)$ & LDoS/QPI map of an isolated defect: $\bbR^2\times\bbR\rightarrow\bbR$ \\
			$\delta\rho(\vec{x},\omega), \delta\rho_0(\vec{x},\omega)$ & local variation of LDoS: $\bbR^2\times\bbR\rightarrow\bbR$ \\
			$n \equiv n_1 \times n_2$ & number of pixels in an $n_1 \times n_2$ measurement/image: $\mathbb{N}$ \\
			$m \equiv m_1 \times m_2$ & number of pixels in an $m_1 \times m_2$ kernel: $\mathbb{N}$ \\
			$s$ & number of observed energy levels (bias voltages): $\mathbb{N}$ \\
			$\mc Y$ & observation of LDoS signature on a finite grid: $\bbR^{n_1\times n_2\times s}$ \\
			$\mc A$ & kernel (from a single defect) on the grid: $\bbR^{m_1\times m_2\times s}$ \\
			$\mc X$ & activation map on the grid: $\bbR^{n_1\times n_2}$ \\
			$\mc Z$ & additive noise \\
			$\boxast$ & convolution operator between each kernel slice and the activation map\\
			Hat: $\hat{\mc A}, \hat{\mc X}, \dots$ & estimates of the corresponding variables \\
			$\Vert\cdot\Vert_F^2$ & Frobenius norm: sum of the squared entries of the given variable\\
			$\mc S$ & the sphere $\{\mc A: \Vert\mc A\Vert_F = 1\}$ \\
			$\psi_\lambda(\mc A, \mc X)$ & objective for fixed $\lambda$ and given $(\mc A, \mc X)$ \\
			$\varphi_\lambda(\mc A)$ & marginalized objective for fixed $\lambda$ and given $\mc A$:  $\varphi_\lambda(\mc A) = \min_{\mathcal{X}} \psi_{\lambda}(\mc A, \mc X)$ \\
			$\eta$ & additive Gaussian noise variance \\
			$\text{SNR}\equiv \frac{var(\mc A)}{\eta}$ & signal-to-noise ratio\\
			$\epsilon(\mc A, \hat{\mc A})$ & error metric: angle between $\mc A$ and $\hat{\mc A}$ on the hemisphere \\
			$\theta$ & kernel concentration: probability that a pixel of $\mc X$ is a defect center
		\end{tabular}
	\end{ruledtabular}
\end{table}

\section{Simulated STM Spectroscopic Measurements on a Square Lattice}
\label{sec:STMSim}

We consider a two-dimensional structure of atoms forming an infinite periodic square array, with
lattice constant $a$. Under the single-electron Tight-Binding (TB) approximation, electrons are
localized to fixed atomic sites~\cite{CMP:AM,CMP:Marder,CMP:EconomouGreen}. This model allows for
an inter-site interaction consisting of a hopping integral between nearest-neighboring lattice
sites, characterized as the fixed hopping-parameter $t<0$. The on-site energies of the lattice are
assumed to be a constant $E_0$. Introducing point defects into the system breaks translational
symmetry of the system and perturbs the Local Density of States (LDoS). These point defects are
assumed to have an energy of $E_j$, where $j$ enumerates the number $N_d$ of defects.

We seek to calculate the LDoS $\rho(\vec{x},\omega)$ for this system with a finite number of point
defects. $\rho(\vec{x},\omega)$ represents the number of states that are available to be occupied by
electrons -- a measure of the relative probability of finding an electron at location $\vec{x}\in
\mathbb{R}^2$ with energy $\omega \in \mathbb{R}$. In STM spectroscopy~\cite{chen2008},
$\rho(\vec{x},\omega)$ is recorded as a function of the probe tip location $\vec{x}$ and an applied
bias voltage $\omega$ between the tip and specimen.

\subsection{Perturbation Theory}

The LDoS is determined by the matrix elements of the Green's Function (GF) in the coordinate representation:
\begin{equation}
\label{eq:defLDoS}
\rho(\vec{x},\omega) = \frac{-1}{\pi} \Im \left[
\bra{\vec{x}}\hat{G}\ket{\vec{x}} \right]
\end{equation}
with the GF defined as $\hat{G}(\omega) = (\omega - \hat{H})^{-1}$.

Assuming no direct coupling between the lattice and point defects, the system Hamiltonian $\hat{H}$
can be expressed as the sum of two contributions $\hat{H} = \hat{H}_0 + \hat{H}_1$ where $\hat{H}_0$
is the TB Hamiltonian for a square lattice and $\hat{H}_1$ is the impurity Hamiltonian:
\begin{equation*}
\hat{H}_1 = \sum_{\alpha=1}^{N_d} E_\alpha \ket{\alpha}\bra{\alpha} 
\end{equation*} 
in which $\alpha$ enumerates the $N_d$ point defects, located at distinct positions
$\vec{x}_\alpha \in \mathbb{R}^2$.

Since $\hat{H}_0$ can be easily diagonalized in $\vec{k}$-momentum space, we treat $\hat{H}_1$ as a
perturbation on $\hat{H}_0$. In accordance with perturbative scattering theory, the GF can be
expressed as:
\begin{equation}
\label{eq:expandG1}
\hat{G} = \hat{G}_0 + \hat{G}_0 \hat{T} \hat{G}_0
\end{equation}
where $\hat{G}_0$ and $\hat{T}$ are the Bare Lattice Green's Function (BLGF) and the scattering
T-matrix, respectively, which
satisfy the following:
\begin{align}
\label{eq:defG0}
\hat{G}_0(\omega) &= \frac{1}{\omega - \hat{H}_0} \\
\label{eq:defT}
\hat{T} &= \hat{H}_1\left(\hat{I} - \hat{G}_0\hat{H}_1 \right)^{-1} 
\end{align}

\subsection{Reduction to Matrix Elements}

Following the prescription in~\eqref{eq:defLDoS}, we compute the matrix elements of $\hat{G}$ from
~\eqref{eq:expandG1} to obtain:
\begin{equation*}
\rho(\vec{x},\omega) = \rho^{(0)}(\vec{x},\omega) + \frac{-1}{\pi} \Im \left[
\bra{\vec{x}}\hat{G}_0\hat{T}\hat{G}_0\ket{\vec{x}} \right]
\end{equation*}
where $\rho^{(0)}(\vec{x},\omega)$ is the LDoS for a system with no defects. Since experimental
probes, such as STM, are sensitive to changes in the LDoS, rather than the LDoS itself, we will
treat $\delta\rho(\vec{x},\omega) \equiv \rho(\vec{x},\omega) - \rho^{(0)}(\vec{x},\omega)$ as the
physical observable:
\begin{equation}
\label{eq:LDoS2} 
\delta \rho(\vec{x},\omega) = \frac{-1}{\pi} \Im \left[
\bra{\vec{x}}\hat{G}_0\hat{T}\hat{G}_0\ket{\vec{x}} \right] =  \frac{-1}{\pi} \Im \left[
\sum_{\alpha,\beta=1}^{N_d} G_0(\vec{x},\vec{x}_\alpha) \, T_{\alpha,\beta} \,
G_0(\vec{x}_\beta,\vec{x}) \right]
\end{equation}
with the following defined quantities:
\begin{align}
\label{eq:defMatGF}
G_0(\vec{x},\vec{y}) &= \bra{\vec{x}}\hat{G}_0\ket{\vec{y}} \\
\label{eq:defMatT}
T_{\alpha,\beta} &= \matrixel{\alpha}{\hat{T}}{\beta}
\end{align}
Henceforth, we will refer to $\delta \rho(\vec{x},\omega)$ in~\eqref{eq:LDoS2} as the LDoS and vice-versa. 

\subsection{Calculation of Matrix Elements}

The following section discusses the calculation of matrix elements of $\hat{G}_0$ and $\hat{T}$ in
~\eqref{eq:defMatGF} and~\eqref{eq:defMatT} necessary to determine $\delta \rho(\vec{x},\omega)$ in
~\eqref{eq:LDoS2}.

\subsubsection{Bare Lattice Green's Function Matrix Elements}
\label{sec:BLGF}

We aim to compute the matrix elements of the BLGF in the coordinate
representation $G_0(\vec{x},\vec{x}';\omega)$. Starting with the definition of the BLGF
in~\eqref{eq:defG0}, we have:
\begin{equation}
\label{eq:BLGF1}
\hat{G}_0 = \int_{\text{BZ}} \diff \vec{k} \frac{1}{\omega - \hat{H}_0} \ket{\vec{k}}
\bra{\vec{k}} = \int_{\text{BZ}} \diff \vec{k} \frac{ 1 }{\omega - E_{\vec{k}}} \ket{\vec{k}}
\bra{\vec{k}}
\end{equation}
where the integration is across the first Brillouin Zone (BZ). The energy dispersion $E_{\vec{k}}$
of the square lattice is:
\begin{equation}
\label{eq:DispersionSquare}
E_{\vec{k}} = E_0 - 2t \left( \cos\left(k_1 a\right) + \cos\left(k_2 a\right) \right)  
\end{equation}
Substituting~\eqref{eq:DispersionSquare} into
~\eqref{eq:BLGF1} gives the BLGF coordinate representation matrix elements:
\begin{align}
\label{eq:BLGF2}
\nonumber \bra{\vec{x}}\hat{G}_0\ket{\vec{x}'} &= \frac{1}{(2\pi)^2} \int_{\text{BZ}} \diff
\vec{k} \, e^{i\vec{k}\cdot \left(\vec{x}- \vec{x}'\right)} \frac{1}{\omega - E_0 - 2t \left(
	\cos\left(k_1 a\right) + \cos\left(k_2 a\right) \right)} \\
G_0(\vec{x},\vec{x}') &= \frac{1}{(2\pi)^2} \frac{1}{2t} \int_{\text{BZ}} \diff
\vec{k} \frac{e^{ik_1(x_1-x'_1)}e^{ik_2(x_2-x'_2)}}{ b - \left( \cos\left(k_1
	a\right) + \cos\left(k_2 a\right) \right) }
\end{align}
where $b \equiv \frac{\omega - E_0}{2t}$ is a dimensionless parameter with a complex $\omega
\rightarrow \omega + i\epsilon$ analytic continuation. Defining the normalized position deviations 
as $s_j \equiv \frac{1}{a}(x_j - x'_j)$,~\eqref{eq:BLGF2} can be reduced to quadratures:

\begin{align}
\label{eq:BLGF3}
\nonumber G_0(\vec{x},\vec{x}') &= \frac{1}{(2\pi)^2} \frac{1}{2t}
\int_{-\frac{\pi}{a}}^{\frac{\pi}{a}}
\int_{-\frac{\pi}{a}}^{\frac{\pi}{a}} \frac{\diff k_1 \diff
	k_2 \,e^{ik_1as_1}e^{ik_2as_2}}{ b - \left( \cos\left(k_1 a\right) + \cos\left(k_2
	a\right) \right) } \\ 
\nonumber G_0(\vec{x},\vec{x}') &= \frac{1}{(2\pi)^2} \frac{1}{2t}\frac{4}{a^2}
\int_{0}^{\pi}\int_{0}^{\pi} \diff \phi_1 \diff \phi_2 \frac{\cos(\phi_1 s_1)\cos(\phi_2
	s_2)}{b - \cos\phi_1 - \cos\phi_2} \\
G_0(\vec{x},\vec{x}') &= \frac{1}{(2\pi)^2} \frac{1}{2t}\frac{4}{a^2}
I_{\text{sq}}\left(\frac{x_1-x'_1}{a},\frac{x_2-x'_2}{a},b\right)
\end{align}
where we define the 2-dimensional definite integral $I_\text{sq}$ as:
\begin{equation}
\label{eq:defIsq}
I_{\text{sq}}(s_1,s_2,b) \equiv \int_{0}^{\pi} \hspace{-4pt} \int_{0}^{\pi} \diff \phi_1 \diff
\phi_2\frac{ \,\cos(s_1\phi_1 )\cos(s_2\phi_2)}{b - \cos\phi_1 - \cos\phi_2}
\end{equation}

\subsubsection{Scattering T-Matrix Elements}
\label{sec:TMat}

We are also interested in computing the matrix elements of $\hat{T}$ between defects, as described
in~\eqref{eq:defMatT}. Recalling that $E_\alpha$ is the on-site energy of the defect located at
$\vec{x}_\alpha$, the defect-defect matrix elements of~\eqref{eq:defT} are:
\begin{align}
\label{eq:TMatrix2}
\nonumber T_{\alpha,\beta} &= \bra{\alpha} \hat{H}_1  \left(\hat{I} - \hat{G}_0 \hat{H}_1
\right)^{-1} \ket{\beta} \\
\nonumber &= E_{\alpha} \bra{\alpha} \left( \hat{I} - \hat{G}_0 \hat{H}_1\right)^{-1}
\ket{\beta} \\
T_{\alpha,\beta}  &=  E_{\alpha} \left( \delta_{\alpha,\beta} - E_{\beta} \,  G_0( \vec{x}_{\alpha}, \vec{x}_{\beta}
) \right)^{-1}
\end{align}
The scattering T-matrix elements are completely determined by~\eqref{eq:TMatrix2} once the BLGF matrix elements are obtained in~\eqref{eq:BLGF3}.

\subsection{Numerical Results on a Single Point Impurity}
Embedded within the $T_{\alpha,\beta}$ calculation in~\eqref{eq:TMatrix2} is a matrix inverse
involving the matrix elements of $\hat{G}_0$ corresponding to inter-defect scatterings. However if
we consider the $N_d = 1$ single impurity limit, we can bypass the formal matrix inversion. In this
limit, the scattering T-matrix becomes a scalar
\begin{equation}
\label{eq:SingleT}
T(\omega) = \frac{1}{ E^{-1}_1 - G_0(\vec{x}_d,\vec{x}_d;\omega) }
\end{equation}
where the impurity location is $\vec{x}_d \in \mathbb{R}^2$. Substituting~\eqref{eq:SingleT} into
~\eqref{eq:LDoS2} gives $\delta \rho(\vec{x},\omega)$ in the presence of a single point impurity:
\begin{align}
\label{eq:SingleLDoS1}
\nonumber \delta \rho(\vec{x},\omega) &= \frac{-1}{\pi} \Im \left[
\frac{G_0(\vec{x},\vec{x}_d;\omega)G_0(\vec{x}_d,\vec{x};\omega)}{E^{-1}_1 -
	G_0(\vec{x}_d,\vec{x}_d;\omega)} \right] \\
\delta \rho(\vec{x},\omega) &= \frac{-1}{\pi} \Im \left[
\frac{G_0^2(\vec{x},\vec{x}_d;\omega)}{E^{-1}_1 -
	G_0(\vec{x}_d,\vec{x}_d;\omega)} \right]
\end{align}

To produce simulated single-defect STM measurements to assess the SBD-STM approach, $\delta
\rho(\vec{x},\omega)$ was computed at all measurement positions $\vec{x}$ for a particular fixed
value of defect energy $E_d$, hopping parameter $t$, measurement energy $\omega$, and specified
defect location $\vec{x}_d = \vec{0}$. Conforming with typical STM  experimental datasets, the
measurement grid consisted of $256 \times 256$ equally-spaced points chosen to overlap with the
square lattice containing $N^2$ atomic sites, with $N = 50$.  $\delta \rho(\vec{x},\omega)$ was
numerically computed at every grid point at 41 different energy $\omega$ values. Typical LDoS maps
of single-defect square lattice systems are shown in Supplementary Figure~\ref{fig:LDoS1}.

\begin{figure*}[ht!]
	\includegraphics[width=0.98\textwidth]{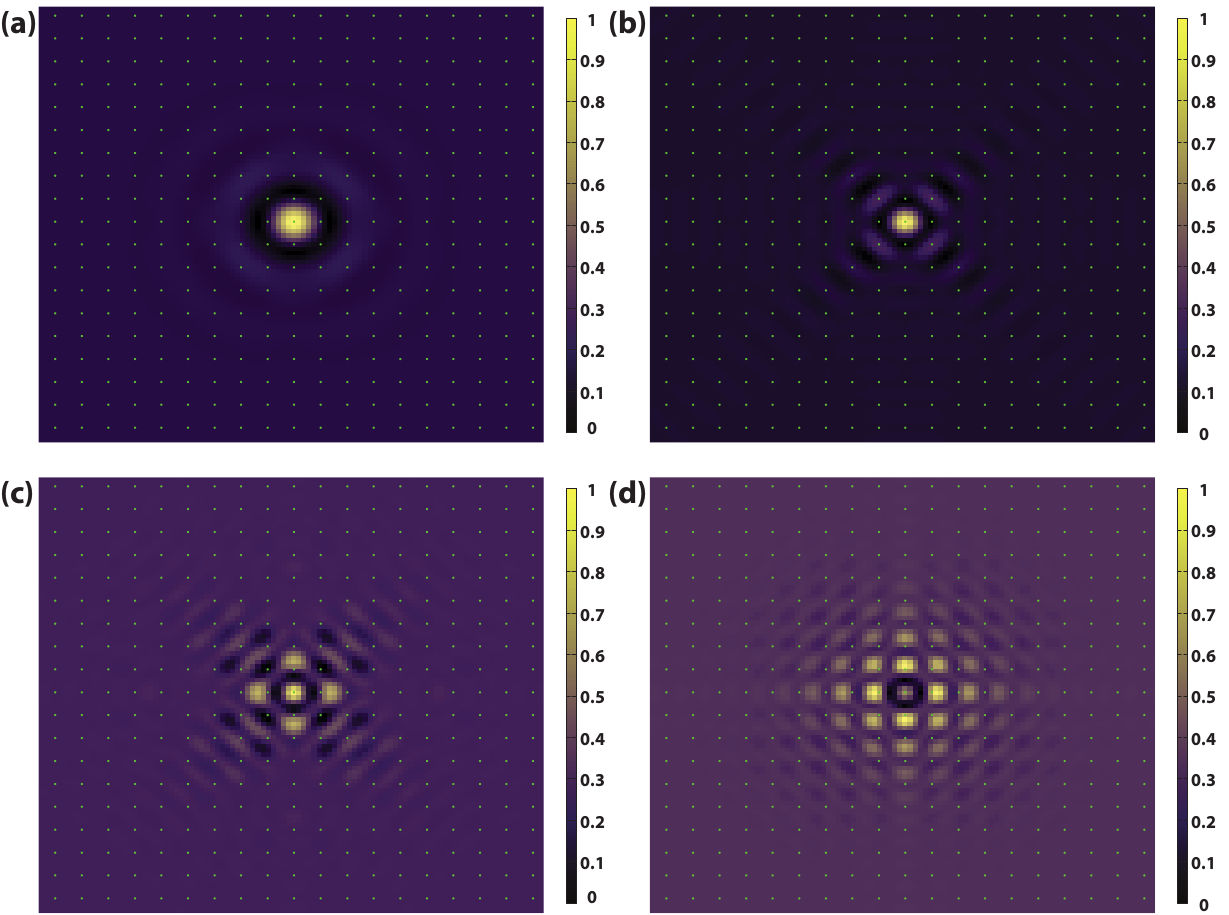}
	\linespread{1}\selectfont{}
	\caption{
		$\delta\rho(\vec{x},\omega)$ for a single impurity on a square lattice with $t = -0.2$ and
		$\omega = -0.5$ (a), $0$ (b), $+0.2$ (c), and $+0.35$ (d). All LDoS maps are normalized between
		0 and 1. The arrays of green dots are guides to the eyes indicating the atomic positions of the
		square lattice.
	}
	\label{fig:LDoS1}
\end{figure*}

One observes that the LDoS resulting from a single impurity has a structure entangled with the
underlying atomic lattice. As the probe energy $\omega$ changes, the LDoS signatures also modulate in
the vicinity of the defect.  Far away from the defect ($\gtrsim 10$ atomic lengths),
$\delta\rho(\vec{x},\omega)$ is nearly constant.

\subsection{Numerical Results on Multiple Point Defects}
Simulations on square lattice systems possessing multiple point defects have also been implemented using~\eqref{eq:LDoS2}. 
Shown in Supplementary Figure~\ref{fig:LDoS2} are representative calculated LDoS maps at different energies for
$70$ point impurities randomly distributed on the square lattice. As expected, the effect of
multiple defect scattering is more vivid at locations where the defects are clustered together. 

\begin{figure*}[ht!]
	\includegraphics[width=0.98\textwidth]{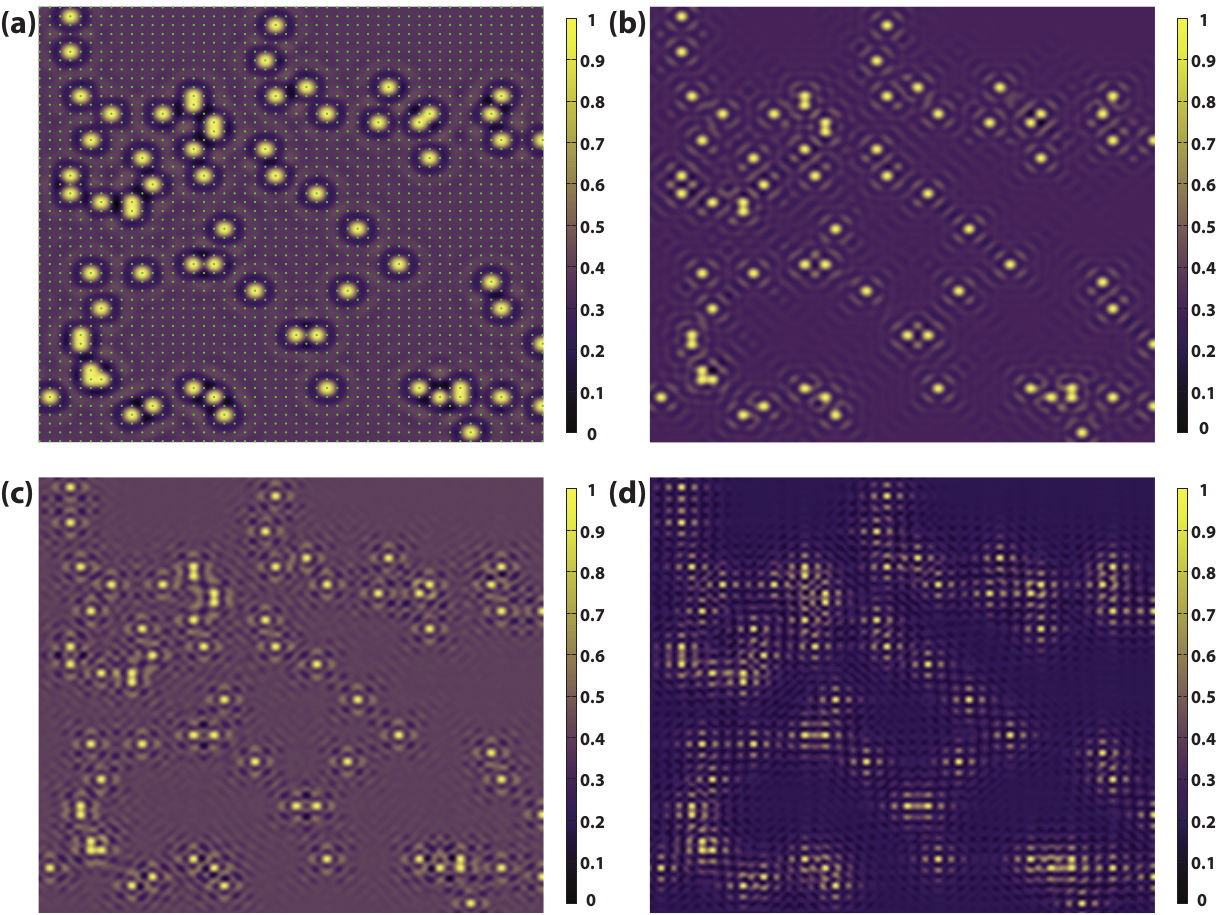}
	\linespread{1}\selectfont{}
	\caption{
		$\delta\rho(\vec{x},\omega)$ for $70$ impurities randomly distributed on a $50 \times 50$ atom
		square lattice with $t = -0.2$ and $\omega = -0.5$ (a), $0$ (b), $+0.2$ (c), and $+0.35$ (d).
		All LDoS maps are normalized between 0 and 1. The array of green dots in (a) is a guide to the
		eyes indicating the atomic positions of the square lattice. Point defects are located at the
		same atomic positions.
	}
	\label{fig:LDoS2}
\end{figure*}

\section{Sparse Inverse Problems} \label{alg_sparse}

The convolutional model proposed in equation (1) of the main text provides a concise description of
STM datasets. However, the problem of inferring $\mc{A}_0$ and $\mc{X}_0$ from $\mc{Y}$ alone is
ill-posed in general -- for $s\in\mathbb{N}$ voltage slices, one must extract $n_1n_2+s\cdot m_1m_2$
values from $s\cdot n_1n_2$ observed pixels. Even with many voltage slices, our kernel slices from
$\mc{A}_0$ are approximately low-pass and vary slowly across bias voltages, so the problem of
recovering $\mc{X}_0$ would remain ill-posed even if $\mc{A}_0$ were known a priori. Since our
problem involves inferring both $\mc{A}_0$ and $\mc{X}_0$, it is essential to incorporate
well-motivated assumptions about the structure of the data.

In this work, we capitalize on the {\em sparsity} of $\mc X_0$ to render the problem well-posed.
Informally, $\mc{X}_0$ is sparse if most of its entries are zero. The sparse signal model is
applicable whenever one is interested in signals that are composed of a few components from a large
``dictionary'': although one can find defects at any location in $\mc{X}_0$ within the convolutional
model, the expected number of defects in typical STM datasets is much smaller than the number of pixels
in the image.

The power of sparsity is most clearly illustrated within the context of {\em linear inverse
	problems}. For instance, suppose one is given a matrix $\mb{D}\in\bbR^{m\times n}$ and an
observation $\mb{y}=\mb{D}\mb{x}_0$, and wishes to recover $\mb{x}_0$. When $m\ll n$, this is an
ill-posed problem with infinitely many solutions. However, if $\mb{x}_0$ is known to be sparse, we
can attempt to recover $\mb{x}_0$ by solving the following optimization problem, which seeks the {\em
	sparsest} solution to the underdetermined linear system $\mb y = \mb D \mb x$:
\begin{equation}
\label{P0}
\begin{cases}
\begin{array}{cc}
\underset{\mb{x}}{\text{minimize}} & \norm{\mb{x}}_0 \\
\text{subject to} & \mb{D}\mb{x} = \mb{y}.
\end{array}
\end{cases}
\end{equation}
Here the $\ell_0$-``norm'' $\norm{\mb{x}}_0$ counts the number of nonzero entries in the vector $\mb
x$. Under very general conditions, the unique optimal solution to~\eqref{P0} is $\mb x_0$, the
signal that generated the observation $\mb y$. For example, for a generic choice of $\mb D$, $\mb x_0$ is the
unique optimal solution to~\eqref{P0}, provided it has nonzero entries, $k$, that are less than half the
number of observations $m  > 2 k$. In this situation, imposing sparsity constraints renders an
ostensibly ill-posed inverse problem well-posed~\cite{donoho2003optimally}.

The optimization in~\eqref{P0} is purely conceptual in nature: it involves a search over all
possible sets of nonzero entries for the vector $\mb x$, and hence is intractable even for small $m$
and $n$. To address this issue, many computationally efficient heuristics have been developed as
alternatives to $\ell_0$-norm minimization for computing sparse solutions. One popular heuristic is
to \emph{relax}~\eqref{P0} by replacing the $\ell_0$-norm $\norm{\mb x}_{0}$ with the $\ell_1$-norm
$\norm{\mb x}_{1} = \sum_{i=1}^n |x_i|$, leading to a convex optimization problem known as {\em
	basis pursuit}:
\begin{equation}
\label{P1}
\begin{cases}
\begin{array}{cc}
\underset{\mb{x}}{\text{minimize}} & \norm{\mb{x}}_1 \\
\text{subject to} & \mb{D}\mb{x} = \mb{y}.
\end{array}
\end{cases}
\end{equation}
This problem can be converted to a linear program, and hence solved in polynomial time using
general-purpose tools. A variety of dedicated algorithms are also available for~\eqref{P1}, which
can efficiently solve very large instances to an accuracy acceptable for modern statistical and
signal processing applications. In short, replacing the $\ell_0$-norm with the $\ell_1$-norm yields
a problem that we can efficiently solve, at scale, on contemporary hardware.

However, a crucial question arises: {\em have we lost anything in moving from the problem that we
	would like to solve ($\ell_0$ minimization) to a problem that we can solve ($\ell_1$ minimization)?}
Specifically, does $\ell_1$ minimization still recover the sparse signal $\mb x_0$? The answer to
the second question is a qualified ``yes.'' Perhaps surprisingly, tractable $\ell_1$ minimizations
also exactly recover $\mb x_0$, provided that (i) $\mb D$ is sufficiently ``nice,'' and
(ii) $\mb x_0$ is sufficiently sparse. For example, for generic (random) $\mb D$, if $\mb x_0$ has at most
$k$ nonzero entries, basis pursuit succeeds when
\begin{equation}
\label{bp_success}
m \geq C \cdot k \cdot\log \left(n/k\right),
\end{equation}
in which $C$ is a fixed constant~\cite{candes2005decoding,vershynin2010introduction}. If we consider $m$
as the number of measurements we have observed, this recovery rate is nearly optimal: $\mb x_0$ has $k$
nonzero entries, and it can be efficiently recovered from about $k \log n$ measurements. These
results hold for random $\mb D$, but deterministic results are also known -- the required property,
coarsely stated, is that sparse vectors $\mb x$ cannot be too close to the nullspace of $\mb D$.
This requirement makes intuitive sense: any sparse vector $\mb x$ that produces a measurement very
close to zero ($\mb y = \mb D \mb x\simeq\mb 0$) will become very difficult to recover, especially
in the presence of noise.

If measurements are contaminated with zero-mean additive noise, i.e. $\mb{y} = \mb{D} \mb{x}_0 +
\mb{w}$, then a popular choice for estimating $\mb{x}_0$ involves solving the LASSO
problem~\cite{hastie2015statistical},
\begin{equation}
\label{P2}
\min_{\mb{x}}
\underbrace{\tfrac{1}{2} \norm{\mb{Dx}-\mb{y}}_2^2}_{\text{data fidelity}} + \lambda \underbrace{\norm{\mb{x}}_1}_{\text{sparsity}}.
\end{equation}
This formulation balances between a standard least squares term, which ensures fidelity to the
observed data, and an $\ell_1$-norm \emph{regularizer}, which biases $\mb x$ towards sparse
solutions. When $\mb{D}$ is randomly chosen, Ref.~\onlinecite{wainwright2009sharp} asserts that --
with similar sampling requirements to~\eqref{bp_success} --  solving~\eqref{P2} will recover the
correct sign pattern from $\mb{x}_0$, provided that $\lambda$ and the entries of $\mb{x}_0$ are
large enough to overcome the bias introduced by the regularizer $\norm{\mb{x}}_1$ and the variance
of the noise term.

Theoretical guarantees of this nature inform the algorithmic design for applications in science and
engineering under a variety of settings; we refer the interested reader to Refs.
~\onlinecite{candes2006near, candes2007sparsity} and the text Ref.
~\onlinecite{foucart2013mathematical}. For example, in biomedical imaging, these efforts help
establish performance and resource requirements in systems where minimal imaging time or radiation
dosage are of critical importance~\cite{lustig2008compressed}.

\subsection{The Sparse Convolutional Model for the STM datasets} 
\label{alg_bd}

In this section, we show how to express the STM model, presented in equation (1) of the main text,
as a convolution between a defect signature and a sparse activation map, and describe how to
estimate the sparse activation map using the LASSO~\eqref{P2}. In this formulation, $\mb D$ plays
the role of the convolution with the defect signature, and $\mb x$ represents the sparse activation
map. Previously, $\mb D$ was assumed to be known beforehand, but in STM data analysis, we must simultaneously
estimate $\mb D$ and $\mb x$. In the next section, we will describe the
technical challenges associated with this harder problem, and offer an efficient algorithm that
accurately estimates both $\mb D$ and $\mb x$, on well-structured numerical examples.

Restricting the STM image and the defect locations $\vecx_j$ to an infinitely large grid of pixel
locations $\bbZ^2$ and using $\gamma$ to denote the Kronecker delta so that $\gamma(\vec{u}) = 1$ if
$\vec{u} = \vec{0}$ and $\gamma({\vecu})=0$ elsewhere, we have:
\begin{equation*}
\drho(\vecx,\omega) = \sum_{j=1}^N c_j \cdot \drho_0(\vecx - \vecx_j,\omega) =   \sum_{\vecu\in\bbZ^2} \drho_0(\vecx - \vecu,\omega) \cdot
\underbrace{\bigg( \sum_{j=1}^N c_j \cdot  \gamma(\vecu - \vecx_j) \bigg)}
_{\text{collect spikes}}.
\end{equation*}
Collecting the defect locations into $\Gamma(\vecx) \equiv \sum_{j=1}^N c_j \gamma(\vecx - \vecx_j)$
leads to the \emph{convolution sum}~\cite{oppenheim2010discrete} between $\drho_0$ and $\Gamma$,
\begin{equation}
\drho(\vecx,\omega) = \sum_{\vecu\in\bbZ^2} \drho_0(\vecx - \vecu,\omega) \cdot \Gamma(\vecu)
= \big( \drho_0 \ast \Gamma \big)(\vecx, \omega).
\label{conv_inf}
\end{equation}

Naturally, mild assumptions are needed on the sizes of the observation $\drho$, defect locations
$\vec{x}_j$, and LDoS signature $\drho_0$. Letting $W_n = \{0,\dots,n_1-1\}\times\{0,\dots,n_2-1\}$
be the observation window, we assume that $\drho(\vecx,\omega) = 0$ for any $\vecx \notin W_n$, and
\ $\bigcup_j\vecx_j \subseteq W_n$, i.e.\ the observations and defect locations are bounded within
an $n_1 \times n_2$ window. We also assume that $m_1 \ll n_1$, and $m_2 \ll n_2$ so that the
individual defect signature takes up a relatively small portion of the observation window $W_m$.

Denoting the discretized versions of the activation map, the LDoS signature of a single
impurity, and the full STM image -- with energies discretized to $s$ levels -- by $\mc{X}_0 \in \bbR^{n_1\times n_2},\
\mc{A}_0\in \bbR^{n_1\times n_2 \times s}$, and $\mc{Y}\in\bbR^{m_1\times m_2 \times s}$
respectively, we have
\begin{align*}
\mc{Y}(\cdot,\omega) &= \mc{A}_0(\cdot,\omega) \ast \mc{X}_0,
\end{align*}
which we express concisely as $\mc{Y} = \mc{A}_0\boxast\mc{X}_0$. $\mc{X}_0$ is expected to be
sparse in this formulation, otherwise the observation would be saturated with defects. Since the
convolution operator $\boxast$ is linear with respect to each argument, a LASSO problem can be
solved to produce a sparse estimate of $\mc{X}_0$ from a noisy observation, provided that $\mc{A}_0$
is known:
\begin{align}
\hat{\mc{X}} \leftarrow
\min_\mc{X} \tfrac{1}{2} \norm{\mc{A}_0 \boxast \mc{X} - \mc{Y}}_F^2 
+ \lambda \norm{\mc{X}}_1
\label{l1_linear}
\end{align}
This is an example of a \emph{Sparse Deconvolution} (SD) problem. (Recall that the \emph{Frobenius
	norm} $\norm{\cdot}_F^2$ is the sum of square entries in the tensorial setting.)

On the contrary, $\mc{A}_0$ is unknown in the STM setting, leading to a bilinear inverse problem
known as \emph{Blind Deconvolution} (BD). However, assuming that a good approximation of the
observation $\mathcal{Y}$ can only be produced by convolving a sparse activation map $\mc X$ with a
candidate kernel $\mc A$ once $\mc{A} \simeq \mc{A}_0$, the optimal objective value from
\eqref{l1_linear} may serve as a basis for finding $\mc{A}$ by formulating the problem
\begin{align}
\big(\hat{\mc{A}}, \hat{\mc{X}}\big) \leftarrow
\min_\mc{A}\min_\mc{X} 
\tfrac{1}{2} \norm{\mc{A} \boxast \mc{X} - \mc{Y}}_F^2 
+ \lambda \norm{\mc{X}}_1
\label{l1_bilinear}
\end{align}
as an instance of the \emph{Sparse Blind Deconvolution} (SBD) problem. This problem is nonconvex,
and consequently, characterizing the performance of efficient algorithms is challenging: currently
available theory does not completely explain the good behavior of simple nonconvex methods on
practical problems. Nevertheless, this problem is part of a rapidly developing area of study with
practical and theoretical implications.

\subsection{Relevant problems and literature}
Recently, variants of the deconvolution and BD problems have attracted significant theoretical
interest. For instance, Refs.~\onlinecite{ahmed2014blind, li2015identifiability, chi2016guaranteed}
study the solvability of BD problems under various settings. However strong assumptions are often
needed, and the settings studied thus far do not staunchly align with our SBD setting~\eqref{l1_bilinear}.

While the difficulty of establishing theory for nonconvex optimization makes guarantees for SBD
problem difficult to formulate, many of the same concerns in existing literature motivate our
SBD-STM algorithm and its analysis: the effect of various sparsity levels, noise power, and the
choice of tradeoff parameter $\lambda$ are motivated by studies involving the LASSO
problem~\eqref{P2}. 

Studies involving variants of the SD problem are more extensive, and provide an important source of
intuition for the SBD problem. For example, Ref.~\onlinecite{candes2014towards} suggests that even
when $\mc{A}_0$ is known in advance, the recoverability of the \emph{activation locations} depends
strongly on the distance between these locations in $\mc{X}_0$ as well as the conditioning of
$\mc{A}_0$ and the noise level. As a result, it is difficult to expect perfect recovery of the
defect locations from $\hat{\mc{X}}$, especially when the kernels in $\mc{A}_0$ are approximately
low-pass and under copious noise.

Practically, deconvolution problems are of interest in a large variety of fields. The extraction of
spike signals in neuroscience~\cite{pnevmatikakis2016simultaneous} and the deblurring of
images~\cite{levin2009understanding} serve as quintessential examples of problems that rely
critically on a (blind) deconvolutional model. We refer interested readers to
Refs.~\onlinecite{chaudhuri2014blind, campisi2016blind} for more details.

\section{Solving the SBD-STM Problem}
\subsection{Symmetries and Nonconvexity}
\label{alg_ncvx}

The bilinearity of the convolutional model $\mc{Y} = \mc{A}_0\boxast\mc{X}_0$ leads to a number of
difficulties characterizing or solving the problem~\eqref{l1_bilinear}. One class of difficulties
arises due to symmetry, in the sense that there are many distinct choices of $(\mc{A}, \mc{X})$ that
are equally sparse \emph{and} approximate $\mc{Y}$ equally well. Having to work with entire
equivalence classes of $(\mc{A}, \mc{X})$ introduces computational issues and makes characterizing
our solution quality of the SBD-STM problem difficult, so modifications for breaking such symmetries
are needed. The methods used to resolve undesirable phenomena arising from such symmetries lead to
many sources of nonconvexity that require special consideration for yielding robust and reliable
estimates of $\mc{A}_0$ and $\mc{X}_0$.


\subsubsection{Optimization on the sphere} 

One consequence of bilinearity is \emph{scaling symmetry}: for any $\mc{A}$, $\mc{X}$, and scalar
$\alpha\in\mathbb{R}$ scaling one variable up and the other down by $\alpha$ leads to the same
convolution, i.e. $\mc{A}\boxast\mc{X}= \big( \alpha\mc{A} \big) \boxast \big( \alpha^{-1} \mc{X}
\big)$. Because the $\ell_1$-norm is a relaxation of the $\ell_0$-norm and not a true sparsity
measure, solutions that essentially have the same data fidelity and sparsity can lead to wildly
different objective values through rescaling by $\alpha$. This undesirable trait is addressed by
fixing $\mc{A}$ to lie on the sphere $\mc{S}=\{\mc{A}\in\bbR^{m_1\times m_2\times
	s}:\norm{\mc{A}}_F=1 \},$ which greatly restricts the equivalence class due to scaling symmetry
to that of sign flips $\mc{A}\boxast\mc{X}= ( -\mc{A} ) \boxast ( - \mc{X} )$. However, the
optimization problem must now be solved over a nonconvex manifold $\mc{S}$, requiring modifications
to standard optimization tools so that operations applied to produce updates are consistent with the
geometry of $\mc{S}$.

\subsubsection{Shifting symmetry produces local minima}
The convolution sum contains a \emph{shifting symmetry} when operating on functions over $\bbZ^2$:
shifting the activation map and the LDoS signature by $\vecDelta=(\Delta_{1},\Delta_{2})$ pixels in
opposite directions yields the same observation. From~\eqref{conv_inf}, observe that
\begin{align*}
\drho(\vecx,\omega) &= \left( \drho_0 \ast \Gamma \right)(\vecx, \omega) \\
&= \sum_{\vecu\in\bbZ^2} \drho_0(\vecx - \vecu,\omega) \cdot \Gamma(\vecu) \\ 
&= \sum_{\vecu\in\bbZ^2} \drho_0 \left(
(\vecx - \vecu) 
+ \vecDelta,\omega \right) 
\cdot \Gamma \left(
\vecu - \vecDelta \right) \\ 
\drho(\vecx,\omega) &=
\mathcal{T}\left[\drho_0;\vecDelta\right](\vecx, \omega)
\ast\ \mathcal{T}\left[\Gamma;-\vecDelta\right](\vecx, \omega)
\end{align*}
where the shift-translation operator $\mathcal{T}$ of a function $f(\vec x)$ is defined as:
$\mathcal{T}\left[f;\vec{\Delta}\right](\vec x) \equiv f(\vec x + \vec{\Delta})$.

The restriction of $\drho$ to the size of $\mc{A}$ breaks this symmetry, since
the shift $\left( \mathcal{T}\left[\mc{A}_0;\vecDelta\right], \mathcal{T}\left[\mc{X}_0;-\vecDelta \right] \right)$ from
$(\mc{A}_0, \mc{X}_0)$ would yield worse data fidelity if any signal from $\drho$ were
shifted outside the size of $\mathcal{A}$. On the other hand, this restriction condition is very weak: the loss in
data fidelity can be insignificant when the window sizes are large. Indeed, such
\emph{shift-truncations} are often achieved by methods employed to solve problems similar to
~\eqref{l1_bilinear} -- we discuss attempts to refine solutions from shift-truncations further in
Section~\ref{alg_imp}.

As a result of the sign and shifting symmetries, objectives related to~\eqref{l1_bilinear} possess
several local minima in the form of \emph{signed shift-truncations}, making the objective
function nonconvex as demonstrated in Supplementary Figure~\ref{fig:fig_hemi}. In general,
nonconvex problems are notorious for local minima or undesired critical points that are
difficult to anticipate and characterize. Indeed, visualizations (see Supplementary Figure
~\ref{fig:fig_hemi}) of the objective in low-dimensional cases confirm that the objective is
geodesically nonconvex in general, containing several saddle points and local minima.

\begin{figure}[ht!]
	\centering \includegraphics[width=.98\textwidth]{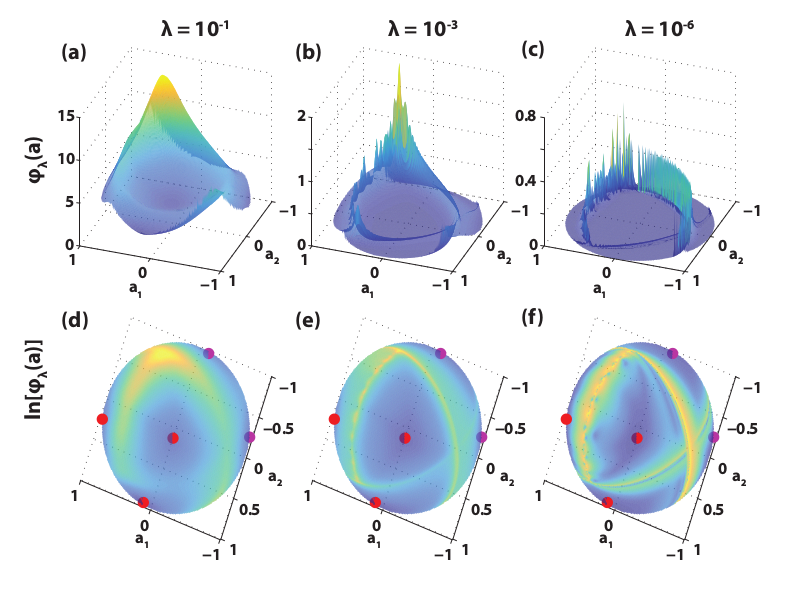}
	\linespread{1}\selectfont{}
	\caption{
		Objective function geometry with varying values of $\lambda$.
		(a)-(c) The objective $\varphi_\lambda(\mb{a}) = \min_\mb{x} \tfrac{1}{2} \norm{\mb{a}\ast\mb{x} -
			\mb{a}_0\ast\mb{x}_0}_2^2 + \lambda \norm{\mb{x}}_1$ over the hemisphere $\left\{(a_1, a_2):
		a_1^2+a_2^2\leq1\right\}$ for $\lambda = 10^{-1}$, $10^{-3}$, and $10^{-6}$. Each pair $(a_1,
		a_2)$ uniquely determines $\mb{a}\in\bbS^2$. The truth kernel is $\mb{a}_0 =
		\mathbb{P}_{\mathbb{S}^2}(\left[1, 8, 2\right])$ (the vector $\left[1, 8,2\right]$ projected onto
		the sphere) and entries of $\mb{x}_0\in\bbR^{256}$ are drawn independently and randomly from a
		Bernoulli-Gaussian distribution with $x_i\overset{\scriptscriptstyle\mrm{i.i.d.}}{\sim}
		\mrm{Bern}(0.1)\cdot\mc{N}(0,1)$. (d)-(f) Logarithm of the objective functions shown in (a)-(c),
		respectively. The truth kernel $\mb{a}_0$ and its shift-truncations
		$\mathbb{P}_{\mathbb{S}^2}(\left[8,2,0\right])$, $\mathbb{P}_{\mathbb{S}^2}(\left[0,1,8\right])$
		are shown in red, and sign-flips $\mathbb{P}_{\mathbb{S}^2}(-\left[8,2,0\right])$,
		$\mathbb{P}_{\mathbb{S}^2}(-\left[0,1,8\right])$ are shown in magenta. Brighter colors on the
		surface plots indicate higher objective value. Notice that each signed shift-truncation shown on
		the hemisphere is close to a corresponding local minimum. Furthermore, the objective landscape
		becomes less regularized as $\lambda$ shrinks, with many local minima appearing for
		$\lambda=10^{-6}$.
	}
	\label{fig:fig_hemi} 
\end{figure}

In such situations, finding a method that can reliably reach local minima becomes more involved. For
the present study we employ the Riemannian Trust-Region Method (RTRM)~\cite{absil2009optimization}.
The standard TRM produces updates for iterates in Euclidean space by minimizing a quadratic
approximation of the objective within a neighborhood of the iterate. The neighborhood radius is
maintained at each iteration to limit the deviation of the quadratic approximation from the true
objective. This procedure results in a second-order method guaranteeing that a local minimum will be
attained so long as the stationary points of the objective are nondegenerate. The RTRM extends this
method to Riemannian submanifolds embedded in Euclidean spaces -- such as the sphere $\mc{S}$ -- by
augmenting TRM operations to become consistent with the manifold geometry. Consequently, the RTRM
provides strong guarantees that a local minimum of the objective will be attained over $\mc{S}$.

Other descent methods can potentially be extended to find local minima in a manifold setting. Except
for pathological examples, solving problems on Euclidean spaces using gradient descent with random
initializations will converge to a local minimum almost surely~\cite{lee2016gradient}, and noisy
gradient descent is guaranteed to efficiently converge to a local minimum~\cite{ge2015escaping}. As
a second-order method, TRM enjoys significantly faster convergence with regards to iterations  but
with increased computation per iteration, as well as better overall tail convergence (when iterates
are close to local minima).

\subsection{Solving the SBD-STM Problem with RTRM}
\subsubsection{Smoothing the Sparse Regularizer} 
\label{alg_sparsereg}

Although we would like to solve problem~\eqref{l1_bilinear} using RTRM, the presence of the
$\ell_1$ regularizer forbids second-order information to be extracted from $\varphi_\lambda(\mc{A})
= \min_{\mathcal{X}}\psi_{\lambda}(\mathcal{X},\mathcal{A}) = \min_\mc{X} \frac{1}{2} \norm{\mc{A}
	\boxast \mc{X} - \mc{Y}}_F^2 + \lambda \cdot r(\mc{X})$. Since the variable $\mc{X}$ is marginalized
by solving the convex problem
\begin{equation}
\label{xstar}
\mc{X}_*(\mc{A}; \lambda) \leftarrow \min_\mc{X} 
\frac{1}{2} \norm{\mc{A} \boxast \mc{X} - \mc{Y}}_F^2 
+ \lambda \cdot r(\mc{X})
\end{equation}
dependent on $\mc{A}$, computing the Hessian $\nabla^2_{\mc{A,A}}\varphi_\lambda(\mc{A})$ 
requires access to $\nabla^2_{\mc{X,X}}\psi_\lambda(\mc{A}, \mc{X})$, which does not exist when
$\ell_1$-norm is present. Instead, a regularizer that enjoys the existence of second derivatives
should be chosen to approximate the $\ell_1$-norm. For this problem, we choose the
\emph{pseudo-Huber} regularizer $r_\mu(\mathcal{X})$, so that:
\begin{equation}
\label{pHuber}
r(\mc{X}) = r_\mu(\mc{X}) \equiv \sum_{i,j} 
\mu^2 \left(\sqrt{1+\mu^{-2} \cdot \mc{X}_{i,j}} - 1\right),
\end{equation}
where $\mu$ would typically be a small positive scalar ($\mu = 10^{-6}$ is chosen in experiments).
We see that although $\underset{\mu\rightarrow 0}{\lim}r(\cdot) = \norm{\cdot}_1$, this function has continuous
derivatives of all orders.

Using a smoothed regularizer also helps to avoid troublesome high-order critical points. Although
the $\ell_1$-norm is a popular regularizer for producing sparse solutions, it can be problematic in
the context of bilinear inverse problems.  When $\mc{A}$ is far from $\pm\mc{A}_0$ on the sphere,
solving $\ \mc{X}_* \leftarrow \min_\mc{X} \frac{1}{2} \norm{\mc{A} \boxast \mc{X} - \mc{Y}}_2^2 +
\lambda \norm{\mc{X}}_1$ encourages entries of $\mc X$ to be exactly zero. If the entire coefficient
map $\mc X$ is zero, then the gradient of $\varphi$ with respect to $\mc A$ is also zero, and
descent methods make no progress. Employing a regularizer that encourages a few elements of
$\mc{X}_*$ to be larg, while keeping other elements small (or ``approximately sparse'') circumvents
this problem as the gradient is unlikely vanish.

\emph{Modified minimization problem.}
Based on the discussion above, we solve the following problem rather than~\eqref{l1_bilinear}:
\begin{equation}
\big(\hat{\mc{A}}, \hat{\mc{X}}\big) \leftarrow
\min_{\mc{A}\in\mc{S}} \left\{ \varphi_\lambda(\mc{A}) \equiv 
\min_\mc{X}  \left[ \psi_\lambda(\mc{A, X}) \equiv
\frac{1}{2} \norm{\mc{A} \boxast \mc{X} - \mc{Y}}_F^2 
+ \lambda \cdot r(\mc{X})\right] \right\}
\label{r_bilinear}
\end{equation}
which is the objective discussed in equation (4) of the main text.

\subsubsection{Implementation of RTRM}
\label{alg_imp}
The MATLAB package ManOpt~\cite{manopt} is used solve~\eqref{r_bilinear} using RTRM. To produce a
quadratic approximation of the objective, one needs the ability to compute $\varphi_\lambda(\mc{A}),
\ \nabla_{\mc{A}}\varphi_\lambda(\mc{A})$ as well as $\nabla^2_{\mc{A,A}}\varphi_\lambda(\mc{A})$
for any $\mc{A}\in\mc{S}$. By rewriting the convolution operator $\boxast$ in terms of \emph{cyclic
	convolutions}, compact expressions can be derived for both the (Euclidean) gradient and Hessian,
which are then mapped to their Riemannian equivalents via tools from the package. As an additional
benefit, expression in terms of cyclic convolutions allows fast computation of $\boxast$ via the
FFT.

As discussed earlier, the $\ell_1$ regularizer is replaced with the pseudo-Huber norm to make
second-order information available to the RTRM. The pseudo-Huber regularizer allows the precise
solution of~\eqref{xstar} to be found using the algorithm proposed by Fountoulakis and
Gondzio~\cite{fountoulakis2016second}, which reports good performance when the observations are
generated from a poorly conditioned linear system. Since the individual voltage slices of
$\mathcal{A}$ are expected to be smooth and low-pass in the Fourier domain, the case where
$(\mc{A}\boxast \cdot)$ is poorly conditioned is of practical interest.

The availability of a method to compute $\varphi_\lambda$ and its Riemannian gradient and Hessian
allows one to solve~\eqref{r_bilinear} using RTRM. We denote this procedure using the notation
\begin{equation*}
\label{alg_ASolve}
\hat{\mc{A}} \leftarrow \mathtt{ASolve}\big(\mc{A}_{\mathrm{init}},\lambda;\mc{Y},(m_{1},m_{2})\big),
\end{equation*}
passing the regularization parameter $\lambda$ and an initialization $\mc{A}_{\mathrm{init}}$ for
$\mc{A}$ as input arguments. The dependence of \texttt{ASolve} on the observation $\mc{Y}$ and the
kernel size $(m_{1},m_{2})$ (which is implicitly provided by $\mc{A}_{\mathrm{init}}$) is explicitly
made, although we will write $\mathtt{ASolve}\big(\mc{A}_{\mathrm{init}},\lambda\big)$ for brevity
when the observation and kernel sizes are clear by context.

\begin{figure}[ht!]
	\centering \includegraphics[width=.98\textwidth]{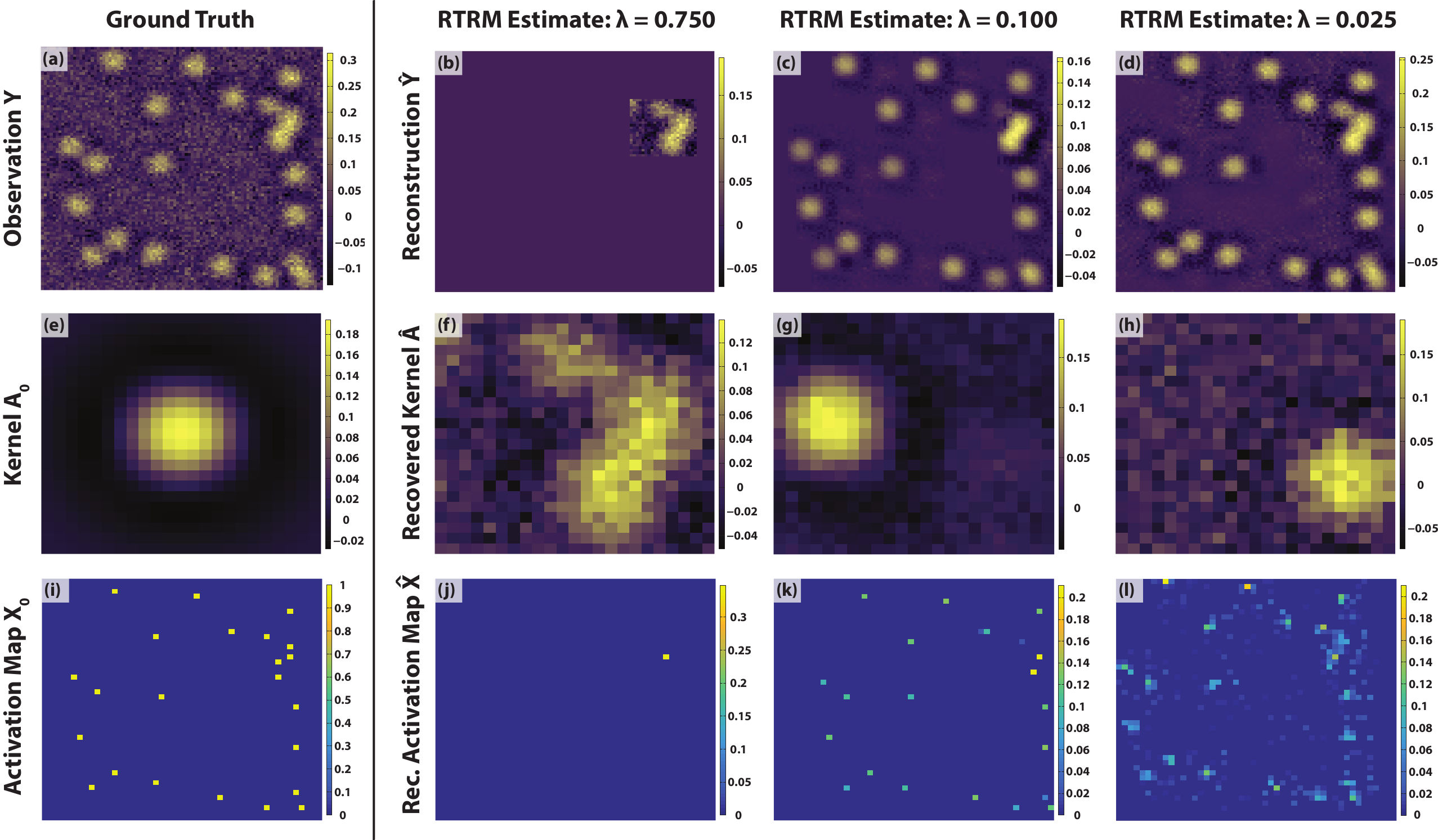}
	\linespread{1}\selectfont{}
	\caption
	{
		RTRM estimates (without refinement) from a simulated observation $\mc{Y} = \mc{A} \ast \mc{X} +
		\mc{Z}$, with moderate noise $(\mc{Z})_{ij}\sim\mc{N}(0,10^{-3})$ and using selected values of
		$\lambda =$ 0.750, 0.100, and 0.025. 
		The original observation $\mathcal{Y}$ is shown in (a), which
		was determined through the convolution of the truth kernel $\mathcal{A}_0$ and truth activation
		map $\mathcal{X}_0$, depicted in (e) and (i), respectively. Reconstructed observations from the
		convolution $\hat{\mathcal{Y}} = \hat{\mathcal{A}} \ast \hat{\mathcal{X}}$ for various values of
		$\lambda$ are shown in (b)-(d). Observe that recovered kernels (f)-(h) $\hat{\mc{A}}$ correspond
		to shift-truncations of the truth kernel $\mathcal{A}_0$ shown in (e), but the low-energy side
		lobes tend to be lost under noise. Recovered activation locations from $\hat{\mc{X}}$ in (j)-(l)
		correspond to shift-truncations of the ground truth $\mc{X}_0$ in (i), but are not exact --
		recovery is particularly poor for nearby activation locations. Recovery in terms of the locations
		and relative magnitudes in $\hat{\mc{X}}$, as well as the details of $\hat{\mc{A}}$, is best when
		$\lambda$ is chosen carefully. Setting $\lambda$ too large causes $\hat{\mc{X}}$ to shrink
		excessively; conversely spurious activation locations are identified when $\lambda$ is too low.
		Both situations degrade the quality of the estimate $\hat{\mc{A}}$.
	}
	\label{fig:shift} 
\end{figure}

Poor estimates of $\hat{\mc{A}}$ can occur due to overfitting the noise term in the observation
$\mc{Y}$. Higher values of $\lambda$ help retain the quality of $\hat{\mc{A}}$ by enforcing sparsity
in $\hat{\mc{X}}$. On the other hand, if $\hat{\mc{X}}$ becomes too sparse ($\lambda$ is too large),
one can fail to recover activation locations from $\mc{X}_0$, meaning that there are fewer defects
from the $\mc{Y}$ being taken into account by \ttt{ASolve}. This can also lead to poor estimates of
$\mc A$.

Empirically, \ttt{ASolve} appears to perform well in simulations in which observations are generated
by ground truths $\mc{A}_0$ and $\mc{X}_0$, and the entries of $\mc{Z}$ are drawn independently from
the Gaussian distribution $\mc{N}(0,\eta)$. In particular, if the noise power $\eta$ is not too high
and $\mc{X}_0$ is sufficiently sparse, then a range of $\lambda\geq0$ will lead $\hat{\mc{A}}$ to
approximate \emph{shift-truncations} of $\mc{A}_0$ regardless of the initialization to \ttt{ASolve}.
This is demonstrated in Supplementary Figure~\ref{fig:shift}.

\subsubsection{The Complete SBD-STM Procedure}
To address the issues of overfitting and shift-truncation, the complete SBD-STM procedure (Algorithm
~\ref{alg_BDSTM}) refines the initial estimate produced by setting $\lambda = \lambda_0$,
\begin{equation*}
\mc{A}_{*}^{(0)} \leftarrow \mathtt{ASolve}\big(\mc{A}_{\mathrm{init}},\lambda_0\big),
\end{equation*}  
through iterations of \texttt{ASolve} over a larger size of $\mathcal{A}$ and a decreasing sequence
$\lambda_{k}$ of regularization parameters until $\lambda_{k}\leq\lambda_{\mrm{end}}$ -- this is
often referred to as a \emph{graduated optimization} or \emph{continuation} procedure.


\begin{figure}[!htb]
	\begin{algorithm}[H] 
		\begin{flushleft}
			\linespread{1}\selectfont{}
			\caption{Complete SBD-STM Procedure}
			\label{alg_BDSTM}
			
			\textbf{Input: }
			\begin{itemize}
				\item Observation \textbf{$\mc{Y}\in\mathbb{R}^{n_{1}\times n_{2}\times s}$},
				\item Kernel size $\left(m_{1},m_{2}\right)$,
				\item Initial $\lambda_{0}\geq0$, decay rate $\alpha\in\left[0,1\right)$,
				and final $\lambda_{\mathrm{end}}\geq0$.
			\end{itemize}
			\textbf{Initial phase:}
			\begin{enumerate}
				\item Randomly initialize: $\mc{A}^{\left(0\right)}\in\mc{S=\mathbb{S}}^{m_{1}\times m_{2}\times s}$.
				\item $\mc{A}_{*}^{\left(0\right)}\leftarrow{\tt ASolve}\left(\mc{A}^{\left(0\right)},\lambda_{0}\right)$.
			\end{enumerate}
			\textbf{Refinement phase:}
			\begin{enumerate}
				\item Lifting: Get $\mc{A}^{\left(1\right)}\in S^{'}=\mc{\mathbb{S}}^{m_{1}^{'}\times m_{2}^{'}\times s}$
				by zero-padding the edges of $\mc{A}_{*}^{\left(0\right)}$ with
				a border of width $\left\lfloor \frac{m_{i}}{2}\right\rfloor$.
				\item Set $\lambda_{1}=\lambda_{0}$.
				\item Continuation: \textbf{Repeat} for $k=1,2,\dots$ \textbf{until} $\lambda_{k}\leq\lambda_{\mathrm{end}}$,
				
				\begin{enumerate}
					\item $\mc{A}_{*}^{\left(k\right)}\leftarrow{\tt ASolve}\left(\mc{A}^{\left(k\right)},\lambda_{k}\right)$,
					\item Centering: 
					
					\begin{enumerate}
						\item Find the size $m_{1}\times m_{2}$ submatrix of $\mc{A}_{*}^{\left(k\right)}$ that maximizes the Frobenius (square) norm across all $m_{1}\times m_{2}$ submatrices.
						\item Get $\mc{A}^{\left(k+1\right)}$ by shifting $\mc{A}_{*}^{\left(k\right)}$
						so that the chosen $m_{1}\times m_{2}$ restriction is in the center,
						removing and zeropadding entries as needed.
						\item Normalize $\mc{A}^{\left(k+1\right)}$ so it lies in $\mc{S}^{'}$.
					\end{enumerate}
					\item Set $\lambda_{k+1}=\alpha\lambda_{k}$.
				\end{enumerate}
			\end{enumerate}
			\textbf{Output:}
			\begin{enumerate}
				\item Extract $\hat{\mc{A}}\in\mc{S}$ by extracting the restriction
				of the final $\mc{A}^{\left(k+1\right)}$ to the center $m_{1}\times m_{2}$
				window.
				\item Find the corresponding activation map $\hat{\mc{X}}\in\mathbb{R}^{n_{1}\times n_{2}}$
				by solving $\min_{\mc{X}}\psi_{\lambda_{k}}\big(\hat{\mc{A}},\mc{X}\big)$.
			\end{enumerate}
		\end{flushleft}
	\end{algorithm}
\end{figure}


To deal with shift-truncations, the kernel size is enlarged to $m^{'}_1 \times m^{'}_2$ in the
refinement phase once the initial estimate $\mc{A}^{(0)}_*$ is obtained. At the end of each
refinement, the variables are shifted so that the $m_1 \times m_2$ submatrix of $\mc{A}^{(k)}_*$
with the largest signal energy is centered. The choice of $m^{'}_1$ and $ m^{'}_2$ depends on how
far one expects the $\mc{A}^{(0)}_*$ to be shifted from the ground truth $\mathcal{A}_0$, but it
cannot be too large to prevent the refinements $\mc{A}^{(k)}_*$ from converging into a different
defect signature far larger than $m_1 \times m_2$ in size. In simulated experiments, we choose
$m_{i}^{'}=m_{i}+2\left\lfloor \frac{m_{i}}{2}\right\rfloor$ for $i=1,2$, as it is unlikely a
shift-truncation by any more than $\frac{m_{i}}{2}$ pixels can be a local minimum of
$\varphi_{\lambda_0}$.

Regarding graduated continuation, we suggest that $\lambda_0$ be chosen relatively large (around 0.1
- 0.5) based our discussion in Section~\ref{alg_imp}. This encourages a sparse $\mc{X}_*(\mc{A};
\lambda)$, and forces the activations to considerably favor regions of the image with prominent
defect signatures. Although $\mc{A}_{*}^{(0)}$ will be affected by heavy bias and noise (see
Supplementary Figure~\ref{fig:shift}), this can be refined in later iterations.

Starting with $\lambda_1=\lambda_0$, a geometrically decreasing sequence for $\{\lambda_k\}$ is
chosen, i.e. $\lambda_k = \alpha\cdot\lambda_{k-1}$ for $k=2,3,\ldots$. This leaves the user to
consider the decay rate $\alpha\in\left[0,1\right)$ and the terminating regularization parameter
$\lambda_{\mrm{end}}$. Based on Ref.~\onlinecite{wainwright2009sharp}, we recommend that
$\lambda_{\mrm{end}}$ be chosen proportionately to $\sqrt{n_1\cdot n_2\cdot \eta}$, in which $\eta$
denotes the additive noise variance. In particular, observe that in the limit as $\lambda\rightarrow
0$ and in the absence of noise,~\eqref{r_bilinear} becomes equivalent to the equally constrained
problem
\begin{align*}
\begin{array}{cc}
\underset{\mc{X}, \mc{A}\in\mc{S}}{\min} & r(\mc{X}) \\
\mathrm{s.t.} & \mc{A}\boxast\mc{X}=\mc{Y}.
\end{array}
\end{align*}
When considering the decay rate $\alpha$, a smaller choice of $\alpha$ means fewer refinements are
needed. However, we would also like to decrease slowly enough so that $\varphi_{\lambda_{k+1}}$ is
not ``too different'' from $\varphi_{\lambda_{k}}$, in the sense that we do not want
$\mc{A}_*^{(k)}$ to jump to a wildly different, possibly malign, local minimum when producing
$\mc{A}_*^{(k+1)}$.

In our benchmarking experiments described in the main text, the varying noise level prohibits us
from carefully tuning $\lambda_{\mrm{end}}$, but even without graduated continuation (single refinement,
$\lambda_{1}=\lambda_{0}=0.5$), the SBD-STM procedure produces qualitatively similar and informative
estimates of $\mc{A}_0$, albeit slightly suboptimal.


\section{Application to Image Deblurring}

In this section, we present an extension of our algorithmic approach that addresses an image
deblurring problem in computer vision. Image deblurring aims to recover a sharp natural image from
its blurred observation due to some unknown photographic process, such as a shaking camera or
defocusing~\cite{Levin2011-PAMI,Kundur1996-SPM,Krishnan2009-NIPS}. Although natural images are
generally not sparse, it is widely acknowledged that sparsity exists in its spatial
gradient~\cite{Levin2011-PAMI,Fergus2006-ACM}.

Suppose $\mb{Y} = \mb{A}_0 \ast \mb{X}_0$ is the observed blurry image, which is represented as the
convolution of the original sharp image $\mb{X}_0$ and a kernel $\mb{A}_0$ that models the blurring.
Owing to the linearity of the convolution operator, the gradient of the observed blurred image must
equal the convolution of the blurring kernel $\mb{A}_0$ and the gradient of the original sharp
image, which possesses the requisite sparsity needed in our algorithm. For instance, we have in two
dimensions:
\begin{align*}
\nabla_x \mb{Y} &= \mb{A}_0 \ast \nabla_x \mb{X}_0 \\
\nabla_y \mb{Y} &= \mb{A}_0 \ast \nabla_y \mb{X}_0
\end{align*}
where $\nabla_x$ and $\nabla_y$ denote derivatives in the $x$ and $y$ directions, respectively. Since the gradients $\nabla_x \mb{X}_0$ and $\nabla_y \mb{X}_0$ are sparse signals by hypothesis, the image deblurring SBD problem can be cast as the following optimization problem:
\begin{equation}
\label{eq:DeblurObjective}
\min_{\mb{A} \in \mathbb{S}_{\ast}, \mb{X}_x, \mb{X}_y} \left[ \frac{1}{2} \norm{ \nabla_x \mb{Y} - \mb{A} \ast \mb{X}_x }_F^2 + \lambda \cdot r(\mb{X}_x) + \frac{1}{2} \norm{ \nabla_y \mb{Y} - \mb{A} \ast \mb{X}_y }_F^2 + \lambda \cdot r  (\mb{X}_y) \right]
\end{equation}
where $\mathbb{S}_{\ast}$ denotes the intersection of the unit sphere and the positive orthant. The optimization variables $\mb{X}_x$ and $\mb{X}_y$ correspond to the horizontal and vertical gradients of $\mb{X}$, respectively.

The non-negativity of the blurring kernel $\mb{A}$ removes sign ambiguity during the recovery
process, contrasted to the STM problem described in Section~\ref{alg_ncvx}. With minor
modifications, the two-stage procedure used for SBD-STM can be applied to image deblurring to
determine reliable estimates of the original sharp image. See Supplementary
Figure~\ref{fig:ImageDeblurring} for a demonstration of the image deblurring process using this
approach.

\begin{figure*}[ht!]
	\includegraphics[width=0.95\textwidth]{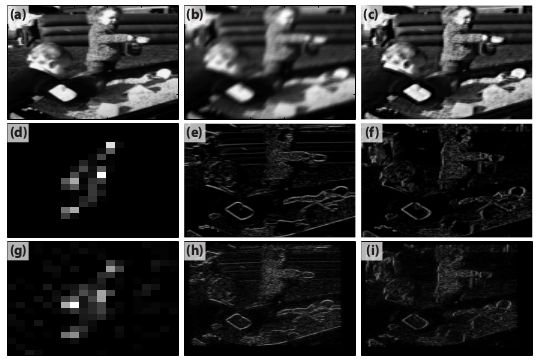}
	\linespread{1}\selectfont{}
	\caption{
		Demonstration of image deblurring by solving the SBD problem in~\eqref{eq:DeblurObjective}. 
		(a) Original sharp image $\mb{X}_0$ (b) Observed blurred image $\mb{Y}$ (c) Recovered deblurred image $\hat{\mb{X}}$ (d) Ground truth of blurring kernel $\mb{A}_0$ (e) Horizontal gradient of original image $\nabla_x\mb{X}_0$ (f) Vertical gradient of original image $\nabla_y\mb{X}_0$ (g) Recovered blurring kernel $\hat{\mb{A}}$ (h) Recovered horizontal gradient $\hat{\mb{X}}_x = \nabla_x \hat{\mb{X}}$ (i) Recovered vertical gradient $\hat{\mb{X}}_y = \nabla_y \hat{\mb{X}}$.
	}
	\label{fig:ImageDeblurring}
\end{figure*}

\section{SBD-STM on High-Dimensional Simulated Measurements}

\begin{figure*}[ht!]
	\includegraphics[width=0.98\textwidth]{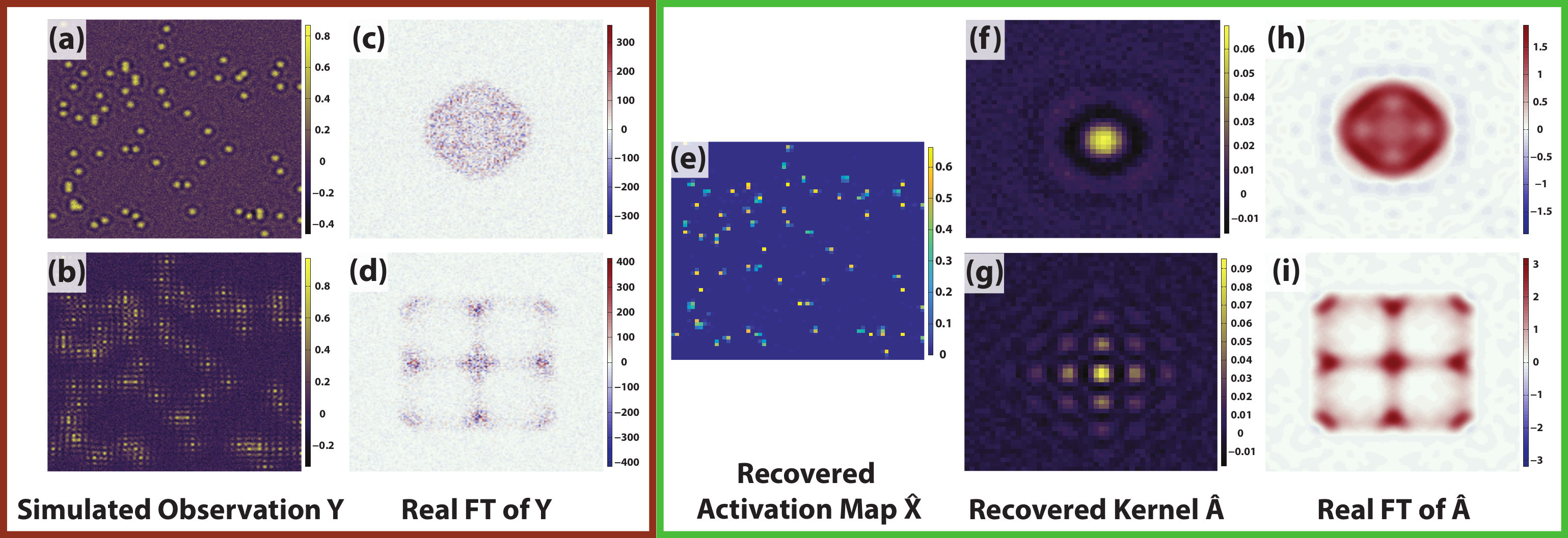}
	\linespread{1}\selectfont{}
	\caption{
		SBD-STM results on simulated noisy STM data with $70$ localized defects. Results for the FT-STM and SBD-STM methodologies on the same dataset are enclosed in the red and green boxes, respectively.
		(a)-(b) Two separate constant-bias slices from the same $\mathcal{Y}\in\mathbb{R}^{256 \times 256
			\times 41}$ dataset (c)-(d) The corresponding Re-FT of the simulated observations in (a)-(b), respectively. Shown in the green box are SBD-STM results: (e)
		the recovered activation map $\hat{\mathcal{X}}$ (globally refined across
		all 41 bias slices) and (f)-(g) the recovered kernel $\hat{\mathcal{A}}$ for each constant-bias slice of
		the observed $\mathcal{Y}$ from (a)-(b). (h)-(i) The corresponding Re-FTs of each recovered kernel $\hat{\mathcal{A}}$. There is significant improvement in data fidelity in the SBD-STM results in (h)-(i) compared to results from the standard FT-STM methodology in (c)-(d). All Re-FT spectra are
		shown with $-3\pi/5a \le k_x,k_y \le 3\pi/5a$.
	}
	\label{fig:STM-SimResults}
\end{figure*}

Using the methods described in Section~\ref{sec:STMSim}, we constructed artifical STM measurements
of a material with 70 identical point defects distributed across a $50 \times 50$ atomic square
lattice. As discussed in the main text, the convolutional data model for STM measurements
has the general form:
\begin{equation*}
\label{eq:ConvolutionalDataModelSTM}
\mathcal{Y} = \mathcal{A} \boxast \mathcal{X} + \mathcal{Z}
\end{equation*} 
The simulated dataset consisted of 41 constant-bias scans of the same $256 \times 256$ pixel
measurement grid and included additive zero-mean Gaussian noise. Supplementary Figure
~\ref{fig:STM-SimResults} displays the algorithm outputs on two separate slices of simulated data.

The Re-FTs of the raw observations $\mathcal{Y}$ show some faint structure, which is obscured by
phase noise associated with random defect locations. By using SBD-STM, one deconvolves the noisy
measurement $\mathcal{Y}$ to recover the underlying kernel $\hat{\mathcal{A}}$ and the activation
map $\hat{\mathcal{X}}$ marking the locations of the defects. As shown on the rightmost column of
Supplementary Figure~\ref{fig:STM-SimResults} the corresponding FTs of the recovered
$\hat{\mathcal{A}}$ are uncontaminated by the phase noise inherent in the FT of $\mathcal{Y}$,
leaving behind the ``true'' structure of the defect. The strong peaks Re-FT spectrum present in
these images are consistent with the allowed elastic scatterings of the underlying square lattice.
Such observations are less evident from Re-FTs of the original $\mathcal{Y}$ since phase noise
suppresses the strong peaks.

\section{Benchmarking the SBD-STM Approach}

We use the artificial STM data in the previous section to generate measurements $\mathcal{Y}$ for
assessing the SBD-STM's ability to provide robust estimates of $\mathcal{A}_0$ and $\mathcal{X}_0$ as
various measurement parameters are varied. Parameters of interest include the measurement size $n
\equiv n_1 \times n_2$, kernel size $m \equiv m_1 \times m_2$, kernel concentration $\theta$, and
additive noise variance $\eta$. To demonstrate the broad applicability of SBD-STM to multiple
modalities of microscopy, we restrict our initial benchmarking datasets to single-energy measurements
$\mathcal{Y}_{\theta,\eta} \in \mathbb{R}^{n_1 \times n_2}$. Under the convolutional data model,
each observation is generated through the process:
\begin{equation}
\label{eq:genSimSTM1}
\mathcal{Y}_{\theta,\eta} = \tilde{ \mathcal{A}_0 } \boxast \mathcal{X}_{\theta} + \mathcal{Z}_{\eta}
\end{equation}
where $\tilde{\mathcal{A}_0} \in \mathbb{R}^{n_1 \times n_2}$ is the zero-padded extension of the
truth kernel $\mathcal{A}_0 \in \mathbb{R}^{m_1 \times m_2}$, which is chosen from TB simulation
results described in Section~\ref{sec:STMSim}. The stochastic contributions for each measurement
$\mathcal{Y}_{\theta,\eta}$ are contained in the random activation map $\mathcal{X}_\theta \in
\mathbb{R}^{m_1 \times m_2}$ as a Bernoulli process with parameter $\theta$ and in $\mathcal{Z}_\eta
\in \mathbb{R}^{m_1 \times m_2}$ as zero-mean Gaussian noise process with variance $\eta$.

A series of independent measurements is produced following~\eqref{eq:genSimSTM1} for several
candidate values of $\theta$ and $\eta$. Each artificial measurement $\mathcal{Y}_{\theta,\eta}$ is
processed by SBD-STM, yielding estimates $\hat{\mathcal{A}}_{\theta,\eta}$ and
$\hat{\mathcal{X}}_{\theta,\eta}$ of $\hat{\mathcal{A}}_0$ and $\hat{\mathcal{X}}_\theta$,
respectively. To reduce experimental ambiguity, the SBD-STM regularizer parameter $\lambda$ is fixed
at 0.5 for these benchmarking trials. In practice, $\lambda$ can be appropriately adjusted to suit
specific measurement and model parameters, leading to results that are generally favorable to
SBD-STM outputs with fixed $\lambda$. As described in the main text, we assess the quality of
SBD-STM kernel recovery by defining the real-space error metric:
\begin{equation*}
\label{eq:defRealError}
\epsilon(\hat{\mathcal{A}}_{\theta,\eta},\mathcal{A}_0 ) \equiv \frac{2}{\pi}\arccos \left| \langle \hat{\mathcal{A}}_{\theta,\eta}, \mathcal{A}_0 \rangle \right| 
\end{equation*}
where $\langle \hat{\mathcal{A}}_{\theta,\eta}, \mathcal{A}_0 \rangle \in [-1,1]$ denotes the inner
product between vectorizations of $\hat{\mathcal{A}}_{\theta,\eta}$ and $\mathcal{A}_0$. The mean of
$\epsilon(\hat{\mathcal{A}}_{\theta,\eta},\mathcal{A}_0 )$ is the average angle (normalized by
$\tfrac{\pi}{2}$) between the two vectors on the hemisphere. Thus this metric provides an indication
of the overall SBD-STM recovery performance for measurements with fixed $\theta$ and $\eta$ while
the spread of $\epsilon(\hat{\mathcal{A}}_{\theta,\eta},\mathcal{A}_0 )$ reveals the stability of
SBD-STM to stochastic fluctuations ascribed to $\theta$ and $\eta$.

\subsection{Real Space Error vs Kernel Concentration and High-Amplitude Noise}

To explore the evolution of $\epsilon(\hat{\mathcal{A}}_{\theta,\eta},\mathcal{A}_0 )$ on the kernel
concentration and a wider range of additive noise variances than in the main text, additional
collections of measurements $\mathcal{Y}_{\theta,\eta}$ of fixed size $n = 185 \times 185$ were
constructed from~\eqref{eq:genSimSTM1} with $\mathcal{A}_0$ of fixed size $m = 35 \times 35$. 20
independent measurements were made for each ordered pair $(\theta,\eta)$, with $\eta$ ranging from
$0$ to $0.01$ and using the same $\theta$ values in Figure 4(b) of the main text. A summary of the
real-space recovery errors for these trials is shown in Supplementary Figure~\ref{fig:Sparsity_Noise_35x35}. 
Supplementary Figure~\ref{fig:Fig_5_Example} shows representative observations $\mathcal{Y}_{\theta,\eta}$ and
the corresponding recovered kernel $\hat{\mathcal{A}}_{\theta,\eta}$ for selected SNR and $\theta$
values in Supplementary Figure~\ref{fig:Sparsity_Noise_35x35}.

\begin{figure*}[ht!]
	\includegraphics[width=0.95\textwidth]{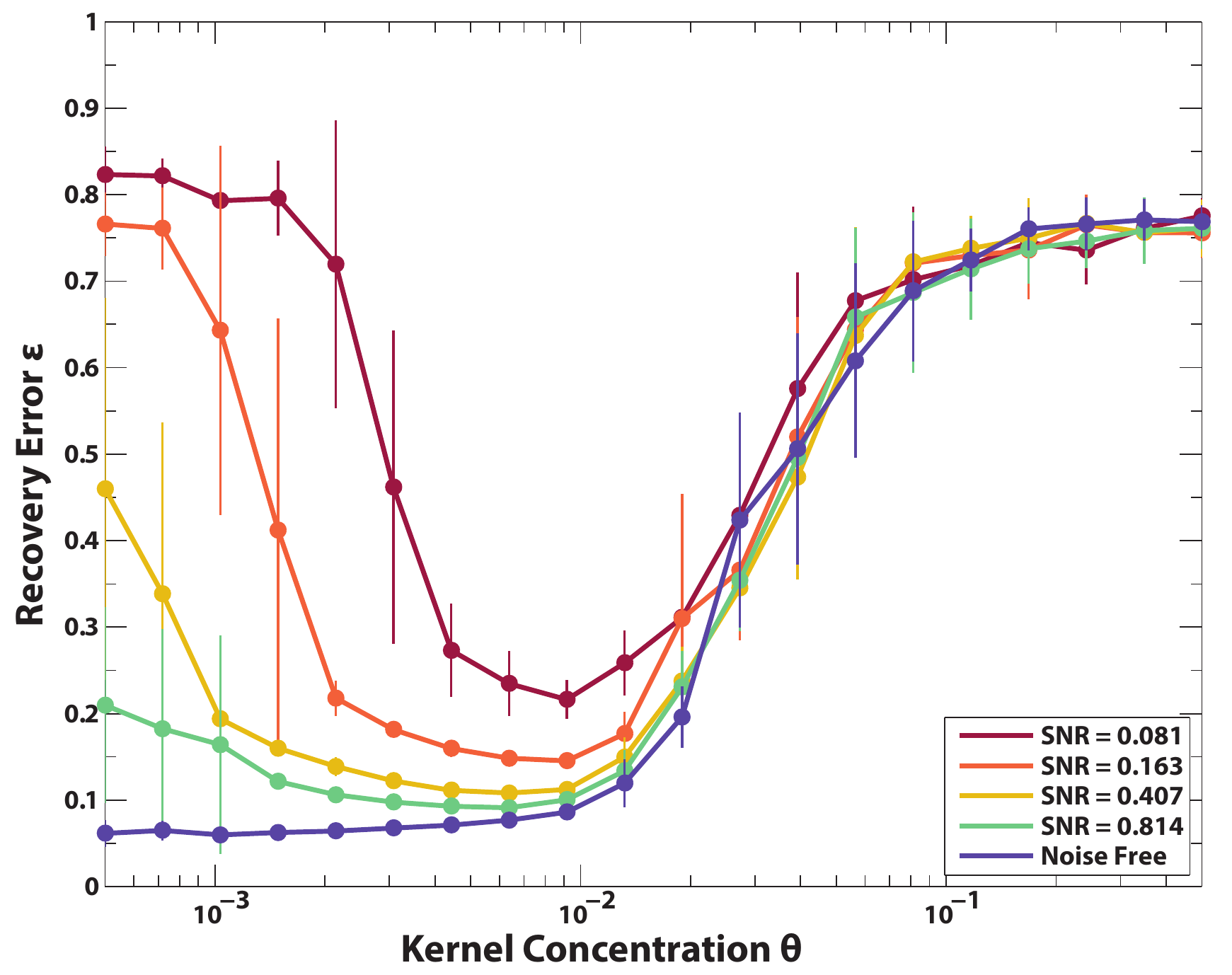}
	\linespread{1}\selectfont{}
	\caption{
		Averaged real-space errors of the recovered kernel $\hat{\mathcal{A}}_{\theta,\eta} \in
		\mathbb{R}^{35 \times 35}$ against the truth kernel $\mathcal{A}_0  \in \mathbb{R}^{35 \times 35}$
		in 20 independent simulated STM measurements $\mathcal{Y}_{\theta,\eta} \in \mathbb{R}^{185 \times
			185}$. Solid lines indicate the kernel recovery error
		$\epsilon(\hat{\mathcal{A}}_{\theta,\eta},\mathcal{A}_0 )$ vs. kernel concentration $\theta$ in
		the presence of additive Gaussian noise with variance $\eta$. Examples of simulated measurements
		in this figure and their corresponding SBD-STM recovered kernels are shown in Supplementary Figure~\ref{fig:Fig_5_Example}.
	}
	\label{fig:Sparsity_Noise_35x35}
\end{figure*}

\begin{figure*}[ht!]
	\includegraphics[width=0.95\textwidth]{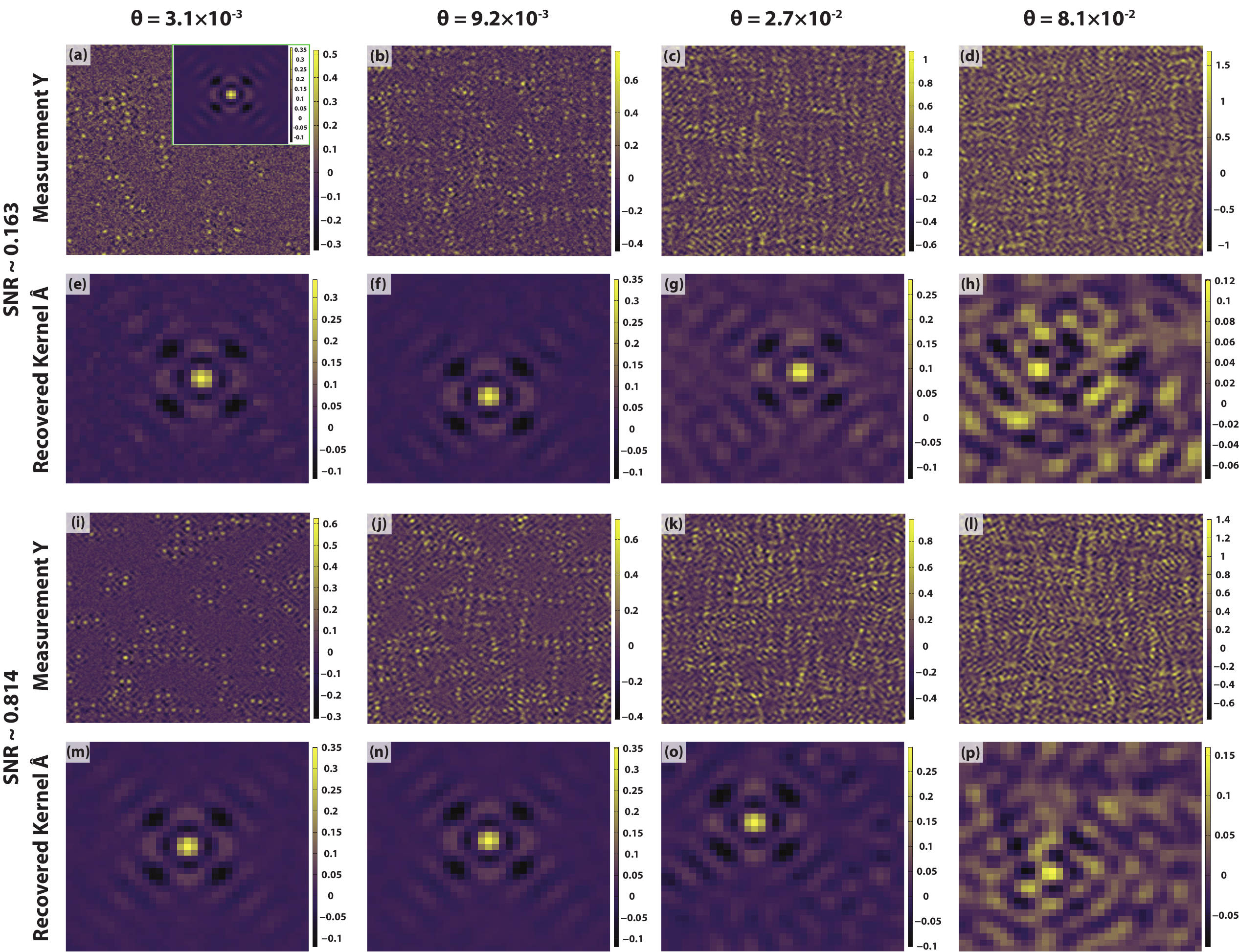}
	\linespread{1}\selectfont{}
	\caption{
		(a)-(p) Example measurements and recovered kernels for select kernel concentrations and SNRs from
		Supplementary Figure~\ref{fig:Sparsity_Noise_35x35}. The truth defect pattern $\mathcal{A}_0$ for these
		measurements is shown in the inset of (a). SBD-STM is able to recover a reasonable estimate of the
		defect kernel $\hat{\mathcal{A}}$, even with poor SNR $\approx 0.163$ that prohibits reliable
		“deconvolution by inspection.” These kernel estimates also include detailed interference patterns
		surrounding the bright peak in the LDoS, which are visually obscured in $\mathcal{Y}$ by the high
		noise levels.
	}
	\label{fig:Fig_5_Example}
\end{figure*}

\subsection{Fourier Space Error Comparison}

Supplementary Figure~\ref{fig:FFT_Example} displays the Real Part of the FTs (Re-FTs) of the real-space results
presented in Supplementary Figure~\ref{fig:Fig_5_Example}. From these examples, one observes that the
Fourier-space errors from raw measurements $\mathcal{Y}_{\theta,\eta}$ gradually improve with
increasing concentration. However, the presence of phase noise and experimental noise washes away
the detailed structure found in the FT of the truth kernel. The corresponding Fourier-space errors
from $\hat{\mathcal{A}}$ exhibit similar trends as the real-space errors with increasing
concentration. There is an improvement in the FT errors as the kernel concentration is increased to
a range spanning $0.001 \lesssim \theta \lesssim 0.01$. However, even in the high concentration
limit (right-most column) the Re-FT of $\hat{\mathcal{A}}$ possesses many detailed features present in
the truth kernel Re-FT that are absent in the Re-FT of $\mathcal{Y}_{\theta,\eta}$.

\begin{figure*}[ht!]
	\includegraphics[width=0.95\textwidth]{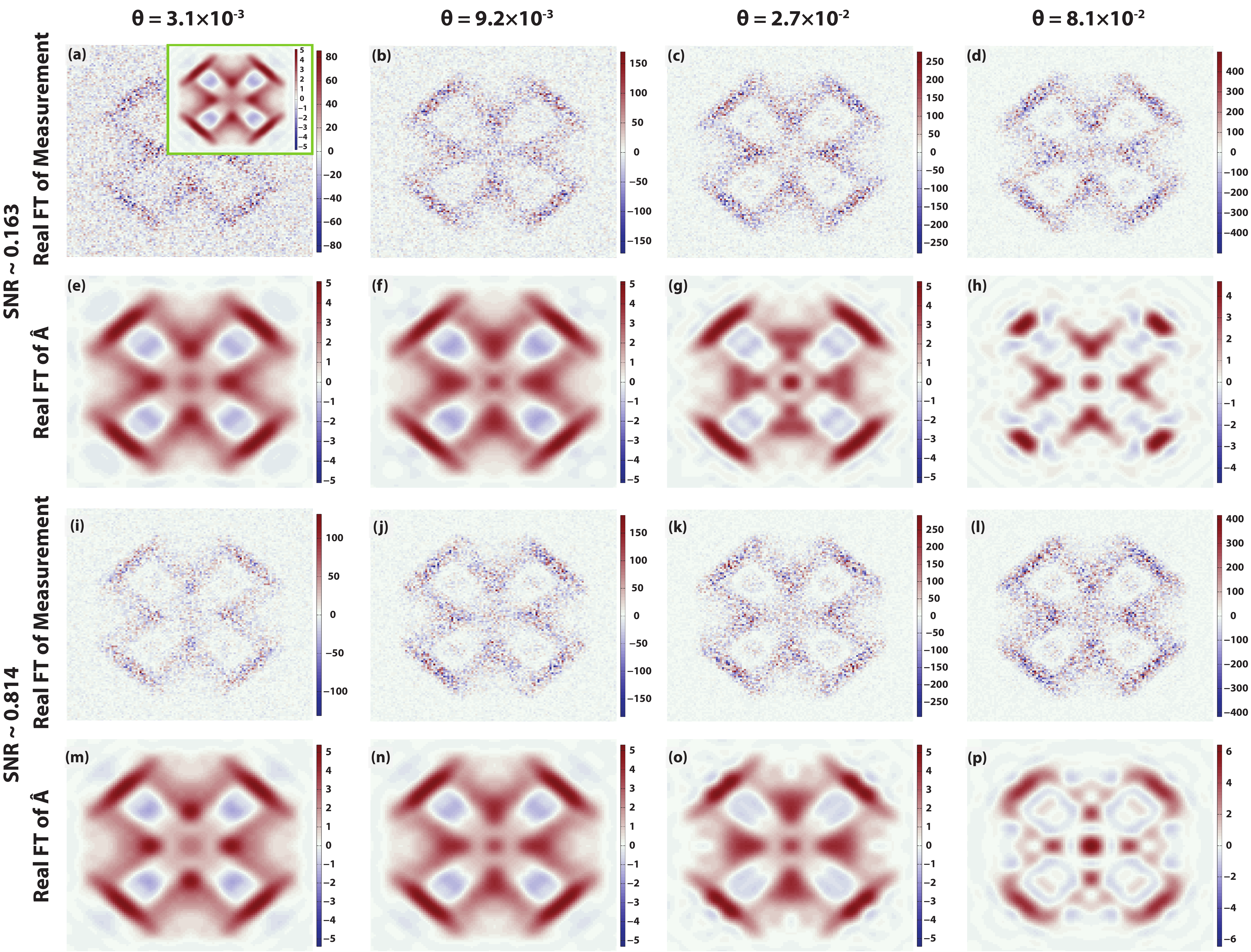}
	\linespread{1}\selectfont{}
	\caption{
		(a)-(p) Re-FTs of the raw observations and recovered kernels from Supplementary Figure
		~\ref{fig:Fig_5_Example}. All Re-FT spectra are shown with $-3\pi/5a \le k_x,k_y \le 3\pi/5a$. The
		Re-FT of the truth defect pattern $\mathcal{A}_0$ for these results is shown in the inset of (a).
		Phase-sensitive recovery of the scattering patterns is only observed in the Re-FT of recovered
		kernels. Note the difference in the colorbar scales between the Re-FT of the original measurements
		and recovered kernels.
	}
	\label{fig:FFT_Example}
\end{figure*}

\bibliography{STM_BD_NatComm_SI_Rev_v2}